\documentclass[twocolumn,showpacs,preprintnumbers,amsmath,amssymb,superscriptaddress,balancelastpage,prb]{revtex4}
\pdfoutput=1

\include{options}
\usepackage {epsfig,color}
\usepackage{graphicx}

\begin{document}

\title{Temperature dependence of the polaronic band structure in strongly correlated electron systems with strong electron-phonon interaction}

\author{I.~A.~Makarov}
\author{S.~G.~Ovchinnikov}
	\email{sgo@iph.krasn.ru}
\affiliation{Kirensky Institute of Physics, Federal Research Center KSC SB RAS, 660036 Krasnoyarsk, Russia}

\date{\today}


\begin{abstract}

In this work we investigate temperature dependence of electronic structure of system with strong electronic correlations and strong electron-phonon interaction modeling cuprates in the frameworks of the three-band p-d-Holstein model by a polaronic version of the generalized tight binding (GTB) method. Within this approach the electronic structure is formed by polaronic quasiparticles constructed as excitations between initial and final polaronic multielectron states. Temperature effect is taken into account by occupation numbers of local excited polaronic states and variations in the magnitude of spin-spin correlation functions and it is manifested in the redistribution of the spectral weight over the Hubbard polaron subbands, and the temperature dependent band structure. Temperature increasing leads to broadening of the spectral function peak at the top of the valence band, shift of the peak, the decreasing of the peak intensity, the growth of bandwidth and reduction of the insulator energy gap. These effects are in a qualitative agreement to the temperature dependence of the ARPES spectra observed in the undoped HTSC cuprates .~\cite{Kim2002,ShenKM2007}

\end{abstract}



\maketitle

\section{Introduction \label{sec_introduction}}
There is still the unsolved question of how electronic structure of cuprates in the normal phase is formed from the undoped La$_2$CuO$_4$. Besides the strong electron correlation (SEC) that results in the Mott-Hubbard insulator ground state of La$_2$CuO$_4$, the undoped copper oxides may have also strong electron-phonon interaction (EPI). The experimental indications for the strong EPI have been found by the ARPES measurements in Ca$_2$CuO$_2$Cl$_2$ that have revealed unusually large linewidth ($\sim1$ eV) in comparison to the narrow line in isostructural Sr$_2$RuO$_4$.~\cite{ShenKM2004} Width of the peak grows with temperature increasing, wherein its position is shifted deep into the band and its spectral weight is lowered.~\cite{Kim2002} Full width at half maximum (FWHM) almost doubles with increasing temperature from $200$ to $400$ K.~\cite{ShenKM2007} Spectra in the optical experiments also demonstrates large temperature effect.~\cite{Onose1999} Explanation of the origin of such experimental features can serve as a key to understanding the nature of quasiparticles forming the electronic structure of cuprates. From the general point of view the study of both effects of SEC and strong EPI may be useful to understand the formation of the normal and superconducting phases in cuprates.

Such mechanisms of temperature influence on electronic structure as thermal broadening and Fermi-Dirac smearing don't reproduce magnitude of effect which is observed in ARPES experiments in the undoped cuprates.~\cite{Kim2002} The most likely reason of large temperature dependence of the ARPES spectra as well as large width of the peak, its Gaussian shape, dispersion of these characteristics is the strong coupling between electrons and phonons and the polaronic origin of Fermi-type excitations in these compounds.~\cite{ShenKM2004} The broad peak of spectral line is considered as the envelope of a number of coherent peaks of multiphonon excitations between local polaron states. Such mechanism of broadening based on Franck-Condon principle which was first formulated for absorption spectra of molecules~\cite{Franck1926,Condon1926,Condon1928} was suggested in Ref.~\onlinecite{ShenKM2004} for cuprates. According to this principle quasiparticles are excitations between vibrational levels of initial and final multielectron states. Photon absorption results in the instantaneous transition of system from $N$-electron state with $\nu $ phonons to $N-1$-electron state with $\nu '$ phonons. Intensity of such transition will be nonzero if equilibrium position of oscillating system of atoms in the final state is shifted relative to initial state. Therefore single peak which would take place for pure electronic transition transforms to several peaks of multiphonon excitations, intensity of every peak depends on Franck-Condon factor (overlap of phonon wave functions of the initial and final states). Temperature dependence is caused by additional thermal population of excited vibrational levels in a Bose-Einstein distribution.~\cite{ShenKM2007} The more general scenario of temperature influence on the electronic structure of systems with many-body effects was suggested in Ref.~\onlinecite{Kim2002}.

Within several theoretical approaches it was confirmed that the key factor for description of ARPES spectra in the undoped cuprates is constructive interplay between magnetic and strong EPI.~\cite{Mishchenko2004,Rosch2005} Coupling with magnetic subsystem gives the correct dispersion and strong interaction with lattice results in the large linewidth of spectra. The theoretical description of the temperature dependence of ARPES spectra also has been carried out by approaches that take into account coupling of carriers both with magnons and phonons. In Ref.~\onlinecite{CatFilMisNag} a growth of the linewidth and binding energy and also spectral weight damping with temperature increasing was obtained within the $t-J$-Holstein model in the frameworks of Hybrid Dynamical Momentum Average self-consistent method. Study of the $t-J$-Holstein model within adiabatic approximation~\cite{RoschGunEurPhysJ} in Ref.~\onlinecite{Rosch2005} also results in the temperature dependence of linewidth. The features of optical conductivity spectra have been qualitatively described with taking into account interaction between hole and lattice and spin degrees of freedom.~\cite{MihFos1990,FalckLevy1993,LupiMaselli1999,Quemerais2000,Fratini2002,Tempere2001} The dynamical mean-field theory for the Holstein $t-J$-model within the polaronic regime qualitatively correctly reproduces filling of a pseudogap with increasing temperature~\cite{Cappelluti2007} which was observed in Ref.~\onlinecite{Onose2004} studying doping and temperature influence on the optical conductivity spectra of the electron-doped Nd$_{2-x}$Ce$_x$CuO$_4$.

The $t-J$ model is known to be the effective low-energy model for cuprates.~\cite{Bulaevskii1968,Chao1977,Hirsch1985} The multiphonon excitations results in the electron states beyond the low-energy window. That is why more general approach in the framework of the multiband $p-d$ model with SEC and strong EPI is desirable.

In this work we investigate the temperature dependence of the electronic structure for the three-band $p-d$-Holstein model by polaronic Generalized Tight-Binding (p-GTB) method.~\cite{Makarov2015} The system under study is compound with SEC and strong EPI the prototype of which is the undoped single-layer cuprate La$_2$CuO$_4$. In the frameworks of GTB method the electronic structure is formed by bands of the Hubbard fermion quasiparticles which are electron excitations between certain initial and final local multielectron states. Spectral weight of excitations depends on probability of transition from initial to final state and filling of these states. Structure and energy of the multielectron states have been found by exact diagonalization of the local part of the total Hamiltonian for the CuO$_6$ cluster with taking into account all Coulomb and electron-phonon interactions in the cluster. With the Hubbard $X$-operators constructed using the exact local multielectron and multiphonon eigenstates the intercluster hopping can be treated in the polaronic version of GTB. It results in the band structure of the Hubbard polarons that are formed by a hybridization of the Hubbard fermions without EPI and the local Franck-Condon resonances.

Since electronic, spin and phonon subsystems are apparently strongly coupled in the HTSC cuprates we should consider the influence of temperature on each of these subsystems in order to obtain the temperature effect on the electronic structure. Direct effect of temperature increasing on electronic subsystem is the occupation of local excited multielectron states with different orbital structure. Hubbard fermions determined by transitions involving these excited states obtain spectral weight which is proportional to the corresponding occupation numbers. Temperature influence on spin system is carried out by the change of magnitude of spin-spin correlations for a short-range order magnetic state (or alternatively by magnetization in the antiferromagnetic state). The third channel is caused by growth of phonon number in the system with temperature. In general the GTB scheme allows taking into account all three channel of the temperature effects. In this paper we have restricted ourselves to the undoped cuprates. The simultaneous treatment of the doping and temperature effects on the electronic structure in systems with SEC and strong EPI will be a subject of a forthcoming paper.

The organization of this work is as follows. In Section~\ref{Model_and_GTB} we briefly remind a scheme of polaronic GTB method and its realization for three-band-Holstein model. Section~\ref{T_dep_without_EPI} devoted to temperature effects on the electronic spectra of the system without EPI in the limit of very strong electron correlations. In Section~\ref{T_dep_with_EPI} the dependence of electronic structure on temperature for system with EPI is discussed. Section~\ref{T_dep_with_EPI_real_par} contains the polaronic band structure and its temperature dependence at the Coulomb parameters typical for La$_2$CuO$_4$. Section~\ref{Summary} is devoted to the main conclusions.

\section{Three-band $p-d$-Holstein model within polaronic GTB-method at finite temperature\label{Model_and_GTB}}

The GTB method~\cite{Ovchinnikov89,Gav2000,LDA+GTB} to calculate the quasiparticle band structure of strongly correlated materials is a version of the cluster perturbation theory in the terms of Hubbard operators. Crystal lattice of atoms is covered by a set of unit cells (separate clusters). The total Hamiltonian of the crystal is divided into two parts. One part is sum of Hamiltonians of separate clusters. Second part is Hamiltonian of intercluster hopping and interactions. The exact diagonalization of the Hamiltonian for a separate cluster with different number of fermions gives the multielectron eigenstates and eigenvalues of the cluster. At this stage all interactions inside the cluster are taken into account. Further Fermi-type excitations between initial and final multielectron states with $N$ and $N+1$ fermions described by the Hubbard operators are introduced. At the last stage of the GTB-method the total Hamiltonian including intercluster interactions is rewritten in the terms of Hubbard operators. Therefore the Hamiltonian of the original model is exactly converted to the multiband Hubbard model in the GTB-method.

It is widely accepted that a low-energy electronic structure of HTSC cuprates is formed by distribution of holes in the CuO$_2$ plane over copper ${d_{{x^2} - {y^2}}}$ (hereinafter $d$) and oxygen ${p_{x,y}}$ orbitals. Phonon system will be described by dispersionless local vibrations of breathing mode, light O atoms vibrate relative to the fixed heavy Cu atoms. Displacement of O atoms from its equilibrium positions changes value of the crystal field for the hole on Cu atom and the hopping integral between copper and oxygen atomic orbitals. In general the EPI includes a diagonal part which is written as renormalization of on-site energy on Cu atom and an off-diagonal EPI which is defined by renormalization of the $p-d$ hopping integral. Evolution of the structure of the local polaronic states and band structure of polaronic quasiparticles with varying diagonal and off-diagonal EPI was discussed in Refs.~\onlinecite{Makarov2015,Makarov2016}. Ratio between diagonal and off-diagonal EPI influences only on character of transition, smooth or abrupt, between states with weak and strong localization (a measure of localization is the average number of holes on $d$-orbital) with varying EPI parameters. In this work we consider only the diagonal EPI. Thus we will use the three-band $p-d$-Holstein model:
\begin{eqnarray}
\label{pd_Hamiltonian}
H & = & {H_{el}} + {H_{ph}} + {H_{e - ph}} \nonumber \\
{H_{el}} & = & \sum\limits_{\bf{f}\sigma } {{\varepsilon _d}d_{\bf{f}\sigma }^\dag {d_{\bf{f}\sigma }}}  + \sum\limits_{\alpha \bf{h}\sigma } {{\varepsilon _p}p_{\alpha\bf{h}\sigma }^\dag {p_{\alpha \bf{h}\sigma }}}  + \nonumber \\
& & + \sum\limits_{\bf{fh}\alpha \sigma } {{{\left( { - 1} \right)}^{{R_{\bf{h}}}}}{t_{pd}}\left( {d_{\bf{f}\sigma }^ \dag {p_{\alpha\bf{h}\sigma }} + h.c.} \right)}  + \nonumber \\
& & + \sum\limits_{\alpha \alpha '\bf{h} \ne \bf{h'}\sigma } {{{\left( { - 1} \right)}^{{M_{\bf{hh'}}}}}{t_{pp}}\left( {p_{\alpha\bf{h}\sigma }^\dag {p_{{\alpha '}\bf{h'}\sigma }} + h.c.} \right)}  + \nonumber \\
& & + \sum\limits_{\bf{f}} {{U_d}d_{\bf{f} \uparrow }^\dag {d_{\bf{f} \uparrow }}d_{\bf{f} \downarrow }^\dag {d_{\bf{f} \downarrow }}}  + \nonumber \\
& & + \sum\limits_{\bf{h}} {{U_p}p_{\alpha\bf{h} \uparrow }^\dag {p_{\alpha\bf{h} \uparrow }}p_{\alpha\bf{h} \downarrow }^\dag {p_{\alpha\bf{h} \downarrow }}}  + \nonumber \\
& & + \sum\limits_{\alpha\bf{fh}\sigma \sigma '} {{V_{pd}}d_{\bf{f}\sigma }^\dag {d_{\bf{f}\sigma }}p_{\alpha\bf{h}\sigma '}^\dag {p_{\alpha\bf{h}\sigma '}}} \nonumber \\
{H_{ph}} & = &\frac{M}{2}\sum\limits_{\bf{h}} {\left( {\dot u_{\bf{h}}^2 + \omega _b^2\dot u_{\bf{h}}^2} \right)} \nonumber \\
{H_{e - ph}} & = & \sum\limits_{\bf{f}\sigma } {{\left( {\sum\limits_{\bf{h}} {{{\left( { - 1} \right)}^{{S_{\bf{h}}}}}{g_d}{u_{\bf{h}}}} } \right)} d_{\bf{f}\sigma }^\dag {d_{\bf{f}\sigma }}}
\end{eqnarray}
Here ${d_{\bf{f}\sigma }}$ and ${p_{\alpha\bf{h}\sigma }}$ are the operators of hole annihilation with spin $\sigma $ on $d$-orbital of the copper atom ${\bf{f}}$ and ${p_x}$(${p_y}$) -orbital of the oxygen atom ${\bf{h}}$, respectively. ${\bf{h}}$ runs over two of the four positions of planar oxygen atoms neighboring to Cu atom in octahedral unit cell centered on site ${\bf{f}}$ at each $\alpha$, ${\bf{h}} = \left\{ {\left( {{f_x} - {a \mathord{\left/
 {\vphantom {a 2}} \right.
 \kern-\nulldelimiterspace} 2},{f_y}} \right),\left( {{f_x} + {a \mathord{\left/
 {\vphantom {a 2}} \right.
 \kern-\nulldelimiterspace} 2},{f_y}} \right)} \right\}$ if $\alpha=x$ and ${\bf{h}} = \left\{ {\left( {{f_x},{f_y} - {b \mathord{\left/
 {\vphantom {b 2}} \right.
 \kern-\nulldelimiterspace} 2}} \right),\left( {{f_x},{f_y} + {b \mathord{\left/
 {\vphantom {b 2}} \right.
 \kern-\nulldelimiterspace} 2}} \right)} \right\}$ if $\alpha=y$, $a$ and $b$ are the lattice parameters. ${\varepsilon _d}$ is the on-site energy of hole on Cu ion and ${\varepsilon _p}$ is the same on O ion; $t_{pd}$ is the amplitude of nearest-neighbor hopping between $d$-orbitals of Cu ion ${\bf{f}}$ and ${p_{x,y}}$-orbitals of O ion ${\bf{h}}$ in CuO$_2$ plane and $t_{pp}$ is the amplitude of nearest-neighbor hopping between ${p_{x,y}}$-orbitals of the oxygen atoms ${\bf{h}}$ and ${\bf{h'}}$. The phase parameters ${R_{\bf{h}}}$ and ${M_{\bf{hh'}} }$ are determined by phases of overlapping wave functions. ${U_d}$ is the Coulomb interaction of two holes on the same copper atom and ${U_p}$ is the same for oxygen atom, $V_{pd}$ is the intersite Coulomb interaction when one hole is on the copper orbital of ${\bf{f}}$ atom and other hole is on the oxygen orbital of ${\bf{h}}$ atom. ${u_{\bf{h}}} = \sqrt {\frac{\hbar }{{2M{\omega _b}}}} \left( {e_{\alpha\bf{h}}^ \dag  + {e_{\alpha\bf{h}}}} \right)$ is the operator of oxygen atom ${\bf{h}}$ displacement, $M$ is the mass of oxygen atom. $e_{\alpha\bf{h}}^\dag $ is the operator of creation of local phonon with frequency ${\omega _b}$, $\alpha$ denotes direction of atom ${\bf{h}}$ displacement. For oxygen atom ${\bf{h}} = \left\{ {\left( {{f_x} - {a \mathord{\left/
 {\vphantom {a 2}} \right.
 \kern-\nulldelimiterspace} 2},{f_y}} \right),\left( {{f_x} + {a \mathord{\left/
 {\vphantom {a 2}} \right.
 \kern-\nulldelimiterspace} 2},{f_y}} \right)} \right\}$ (${\bf{h}} = \left\{ {\left( {{f_x},{f_y} - {b \mathord{\left/
 {\vphantom {b 2}} \right.
 \kern-\nulldelimiterspace} 2}} \right),\left( {{f_x},{f_y} + {b \mathord{\left/
 {\vphantom {b 2}} \right.
 \kern-\nulldelimiterspace} 2}} \right)} \right\}$) displacement is along $\alpha=x$ ($y$) axis. ${g_d}$ is the parameter of the diagonal EPI between hole, located on copper atom and phonon of breathing mode. The phase parameter ${S_{\bf{h}} } = 0$ for ${\bf{h}} = \left( {{f_x} + {a \mathord{\left/
 {\vphantom {a 2}} \right.
 \kern-\nulldelimiterspace} 2},{f_y}} \right),\left( {{f_x},{f_y} + {b \mathord{\left/
 {\vphantom {b 2}} \right.
 \kern-\nulldelimiterspace} 2}} \right)$ and ${S_{\bf{h}} } = 1$ for ${\bf{h}} = \left( {{f_x} - {a \mathord{\left/
 {\vphantom {a 2}} \right.
 \kern-\nulldelimiterspace} 2},{f_y}} \right),\left( {{f_x},{f_y} - {b \mathord{\left/
 {\vphantom {b 2}} \right.
 \kern-\nulldelimiterspace} 2}} \right)$, it is consistent with modulation of the on-site energy for the breathing mode. We introduce dimensionless EPI parameters ${\lambda _{d}} = {{{{\left( {{g_{d}}\xi } \right)}^2}} \mathord{\left/
 {\vphantom {{{{\left( {{g_{d}}\xi } \right)}^2}} {W\hbar {\omega _b}}}} \right.
 \kern-\nulldelimiterspace} {W\hbar {\omega _b}}}$, where $\xi  = \sqrt {\frac{\hbar }{{2M{\omega _b}}}} $, $W$ is the bandwidth of the free electron in tight-binding method without EPI, we accept here $W = 1$ eV and the breathing mode phonon energy $\hbar {\omega _b} = 0.04$ eV.

Since each planar oxygen atom belongs to the two CuO$_6$ clusters at once we should make orthogonalization procedure. We proceed from the hole atomic oxygen orbitals $\left| {{p_{x{\bf{h}}}}} \right\rangle $ and $\left| {{p_{y{\bf{h}}}}} \right\rangle $ to the molecular oxygen orbitals $\left| {{b_{\bf{f}}}} \right\rangle $ and $\left| {{a_{\bf{f}}}} \right\rangle $ by transformation in the $k$-space:~\cite{Shastry89}
\begin{eqnarray}
\label{Shastry}
{b_{\bf{k}}}&=& \frac{i}{{{\mu _{\bf{k}}}}}\left( {{s_{{\bf{k}}x}}{p_{x{\bf{k}}}} - {s_{{\bf{k}}y}}{p_{y{\bf{k}}}}} \right)\nonumber \\
{a_{\bf{k}}}&=& - \frac{i}{{{\mu _{\bf{k}}}}}\left( {{s_{{\bf{k}}y}}{p_{x{\bf{k}}}} + {s_{{\bf{k}}x}}{p_{y{\bf{k}}}}} \right)
\end{eqnarray}
where ${s_{{\bf{k}}x}} = \sin \left( {{{{k_x}a} \mathord{\left/
 {\vphantom {{{k_x}a} 2}} \right.
 \kern-\nulldelimiterspace} 2}} \right)$, ${s_{{\bf{k}}y}} = \sin \left( {{{{k_y}b} \mathord{\left/
 {\vphantom {{{k_y}b} 2}} \right.
 \kern-\nulldelimiterspace} 2}} \right)$ and ${\mu _{\bf{k}}} = \sqrt {s_{{\bf{k}}x}^2 + s_{{\bf{k}}y}^2} $. States ${b_{\bf{f}}}$(${a_{\bf{f}}}$) in the neighbor CuO$_6$ clusters are orthogonal to each other. Similar transformation of the phonon states is undertaken. States of atomic phonons $e_{\bf{h}}^\dag \left| 0 \right\rangle $ are replaced by "molecular" oxygen phonon wave functions $A_{\bf{f}}^\dag \left| 0 \right\rangle $ and $B_{\bf{f}}^\dag \left| 0 \right\rangle $, where operators ${A_{\bf{k}}}$ and ${B_{\bf{k}}}$ in the momentum space are defined by transformation:

\begin{eqnarray}
\label{PhOper}
{A_{\bf{k}}} =  - \frac{i}{{{\mu _{\bf{k}}}}}\left( {{s_{{\bf{k}}x}}{e_{x{\bf{k}}}} + {s_{{\bf{k}}y}}{e_{y{\bf{k}}}}} \right)\nonumber \\
{B_{\bf{k}}} =  - \frac{i}{{{\mu _{\bf{k}}}}}\left( {{s_{{\bf{k}}y}}{e_{x{\bf{k}}}} - {s_{{\bf{k}}x}}{e_{y{\bf{k}}}}} \right)
\end{eqnarray}
Phonon wave functions $B_{\bf{f}}^\dag \left| 0 \right\rangle $ have negligible contribution to the low-energy local cluster eigenstates and consequently to the electron spectral function of the top of the valence band and the bottom of the conduction band. Therefore we neglect them in the further consideration taking into account only $A$-mode phonons.

After orthogonalization procedure the Hamiltonian of three-band $p-d$-Holstein model can be written as
\begin{eqnarray}
\label{Full_Ham}
& H & = \sum\limits_{{\bf{f}}\sigma } {{\varepsilon _d}d_{{\bf{f}}\sigma }^\dag {d_{{\bf{f}}\sigma }}}  + \sum\limits_{{\bf{f}}\sigma } {{\varepsilon _p}b_{{\bf{f}}\sigma }^\dag {b_{{\bf{f}}\sigma }}}  - \nonumber \\
& & - 2{t_{pd}}\sum\limits_{{\bf{fg}}\sigma } {{\mu _{\bf{fg}}}\left( {d_{{\bf{f}}\sigma }^\dag {b_{{\bf{g}}\sigma }} + h.c.} \right)}  - \nonumber \\
& & - 2{t_{pp}}\sum\limits_{{\bf{fg}}\sigma } {{\nu _{\bf{fg}}}\left( {b_{{\bf{f}}\sigma }^\dag {b_{{\bf{g}}\sigma }} + h.c.} \right)}  +\nonumber \\
& & + \sum\limits_{\bf{f}} {{U_d}d_{{\bf{f}} \uparrow }^\dag {d_{{\bf{f}} \uparrow }}d_{{\bf{f}} \downarrow }^\dag {d_{{\bf{f}} \downarrow }}}  + \nonumber \\
& & + \sum\limits_{\bf{fghl}} {{U_p}{\Psi _{\bf{fghl}}}b_{{\bf{f}} \uparrow }^\dag {b_{{\bf{g}} \uparrow }}b_{{\bf{h}} \downarrow }^\dag {b_{{\bf{l}} \downarrow }}}  + \nonumber \\
& & + \sum\limits_{{\bf{fgh}}\sigma \sigma '} {{V_{pd}}{\Phi _{\bf{fgh}}}d_{{\bf{f}}\sigma }^\dag {d_{{\bf{f}}\sigma }}b_{{\bf{g}}\sigma '}^\dag {b_{{\bf{h}}\sigma '}}}  + \nonumber \\
& & + \sum\limits_{\bf{f}} {\hbar {\omega _b}A_{\bf{f}}^\dag {A_{\bf{f}}}}  + \nonumber \\
& & + \sum\limits_{\bf{fg}} {2{g_d}{\xi _d}{\mu _{\bf{fg}}}\sum\limits_\sigma  {\left( {A_{\bf{f}}^\dag  + {A_{\bf{f}}}} \right)d_{{\bf{g}}\sigma }^\dag {d_{{\bf{g}}\sigma }}} }
\end{eqnarray}
where structural factors ${\mu _{\bf{fg}}}$, ${\nu _{\bf{fg}}}$, ${\rho _{\bf{fgh}}}$ are defined as
\begin{eqnarray}
\label{str_factors}
{\mu _{\bf{fg}}} & = & {1 \mathord{\left/
 {\vphantom {1 N}} \right.
 \kern-\nulldelimiterspace} N}\sum\limits_{\bf{k}} {{\mu _{\bf{k}}}{e^{ - i{\bf{k}}\left( {{\bf{f}} - {\bf{g}}} \right)}}} \nonumber \\
{\nu _{\bf{fg}}} & = & {1 \mathord{\left/
 {\vphantom {1 N}} \right.
 \kern-\nulldelimiterspace} N}\sum\limits_{\bf{k}} {{{\left( {{{2\sin \left( {{{{k_x}} \mathord{\left/
 {\vphantom {{{k_x}} 2}} \right.
 \kern-\nulldelimiterspace} 2}} \right)\sin \left( {{{{k_y}} \mathord{\left/
 {\vphantom {{{k_y}} 2}} \right.
 \kern-\nulldelimiterspace} 2}} \right)} \mathord{\left/
 {\vphantom {{2\sin \left( {{{{k_x}} \mathord{\left/
 {\vphantom {{{k_x}} 2}} \right.
 \kern-\nulldelimiterspace} 2}} \right)\sin \left( {{{{k_y}} \mathord{\left/
 {\vphantom {{{k_y}} 2}} \right.
 \kern-\nulldelimiterspace} 2}} \right)} {{\mu _{\bf{k}}}}}} \right.
 \kern-\nulldelimiterspace} {{\mu _{\bf{k}}}}}} \right)}^2}{e^{ - i{\bf{k}}\left( {{\bf{f}} - {\bf{g}}} \right)}}} \nonumber \\
\rho _{{\bf{fgh}}}^A & = &\rho _{{\bf{f}} - {\bf{g}},{\bf{g}} - {\bf{h}}}^A{{ = 1} \mathord{\left/
 {\vphantom {{ = 1} {{N^2}}}} \right.
 \kern-\nulldelimiterspace} {{N^2}}}\sum\limits_{{\bf{kq}}} {{1 \mathord{\left/
 {\vphantom {1 {{\mu _{\bf{k}}}{\mu _{\bf{q}}}}}} \right.
 \kern-\nulldelimiterspace} {{\mu _{\bf{k}}}{\mu _{\bf{q}}}}}}  \times \nonumber \\
& \times & \left[ {\sin \left( {{{{k_x}} \mathord{\left/
 {\vphantom {{{k_x}} 2}} \right.
 \kern-\nulldelimiterspace} 2}} \right)\sin \left( {{{{q_x}} \mathord{\left/
 {\vphantom {{{q_x}} 2}} \right.
 \kern-\nulldelimiterspace} 2}} \right)\cos \left( {{{\left( {{k_x} + {q_x}} \right)} \mathord{\left/
 {\vphantom {{\left( {{k_x} + {q_x}} \right)} 2}} \right.
 \kern-\nulldelimiterspace} 2}} \right) + } \right.\nonumber \\
& + & \left. {\sin \left( {{{{k_y}} \mathord{\left/
 {\vphantom {{{k_y}} 2}} \right.
 \kern-\nulldelimiterspace} 2}} \right)\sin \left( {{{{q_y}} \mathord{\left/
 {\vphantom {{{q_y}} 2}} \right.
 \kern-\nulldelimiterspace} 2}} \right)\cos \left( {{{\left( {{k_y} + {q_y}} \right)} \mathord{\left/
 {\vphantom {{\left( {{k_y} + {q_y}} \right)} 2}} \right.
 \kern-\nulldelimiterspace} 2}} \right)} \right] \times \nonumber \\
& \times & {e^{ - i{\bf{k}}\left( {{\bf{f}} - {\bf{g}}} \right)}}{e^{ - i{\bf{q}}\left( {{\bf{g}} - {\bf{h}}} \right)}}
\end{eqnarray}
Values of coefficients ${\Psi _{\bf{fghl}}}$ and ${\Phi _{\bf{fgh}}}$ strongly decrease with increasing distance between sites ${\bf{f,g,h}}$ ~\cite{Feiner1996}, therefore we restrict our consideration by only intracluster Coulomb interactions, ${\Psi _{0000}} = 0.2109$, ${\Phi _{000}} = 0.918$. Now we can divide Hamiltonian $H$ on the intracluster part and intercluster interactions:
\begin{equation}
\label{Hc_Hcc_tot}
H = {H_c} + {H_{cc}}, {H_c} = \sum\limits_{\bf{f}} {{H_{\bf{f}}}}, {H_{cc}} = \sum\limits_{\bf{fg}} {{H_{\bf{fg}}}}
\end{equation}

\begin{figure}
\center
\includegraphics[width=1.0\linewidth]{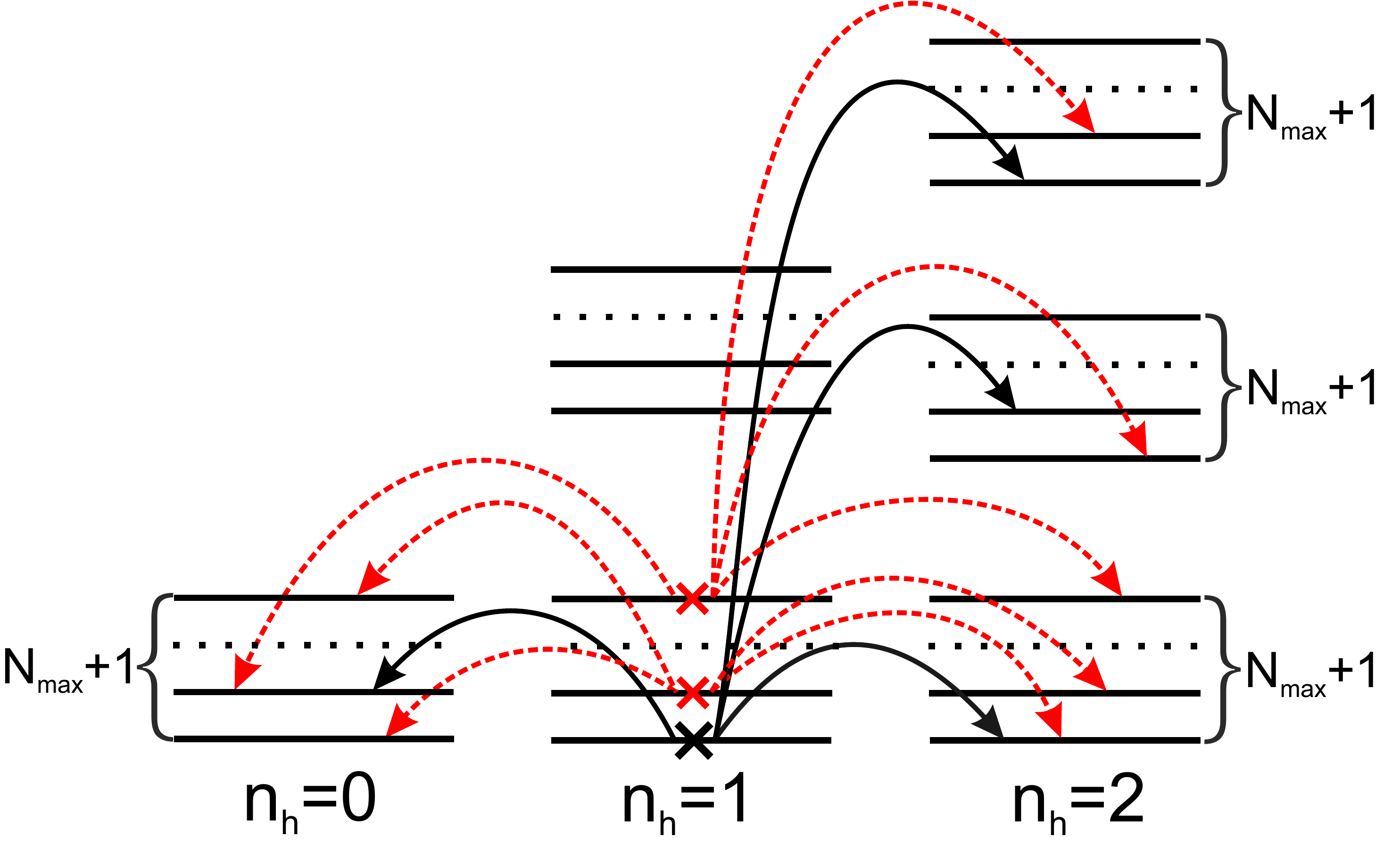}
\caption{\label{fig:levels} Schematic picture of the multielectron vibronic eigenstates of CuO$_6$ cluster (black horizontal lines) with hole numbers ${n_h} = 0,1,2$ and Fermi-type excitations between them. Black solid arrows denotes excitations with nonzero spectral weight at $T = 0$ K due to occupation of single-hole ground state (black cross) for undoped La$_2$CuO$_4$. At nonzero temperature excited single-hole states are thermally occupied (red crosses) and excitations involving these states acquire spectral weight (red dashed arrows).}
\end{figure}

Eigenstates of the CuO$_6$ cluster with hole numbers ${n_h} = 0,1,2$ obtained by exact diagonalization of Hamiltonian ${H_c}$ include hole and phonon basis wave functions. In the hole vacuum sector of the Hilbert space with ${n_h} = 0$ eigenstates are just the harmonic oscillator states:
\begin{equation}
|0,\nu \rangle  = |0\rangle |\nu \rangle, \nu  = 0,1,...,{N_{\max }}
\label{es0}
\end{equation}
Here $|0\rangle $ denotes electronic configuration $\left| {3{d^{10}}2{p^6}} \right\rangle $, and $|\nu \rangle $ - phonon state with phonon number ${n_{ph}} = \nu $. Phonon states $\left| \nu  \right\rangle $ are $\nu $-times action of phonon creation operator ${A^\dag }$ on vacuum state   of harmonic oscillator:
\begin{equation}
\left| \nu  \right\rangle  = \frac{1}{{\sqrt {\nu !} }}{\left( {{A^\dag }} \right)^\nu }\left| {0,0,...,0} \right\rangle
\label{harm_osc}
\end{equation}
 ${N_{\max }}$ is the cutoff for number of phonons, it is calculated for each certain set of parameters from the following condition: addition of phonon number above ${N_{\max }}$, $N > {N_{\max }}$, does not change the electron spectral function. The value ${N_{\max }}$ mainly depends on the EPI coupling parameter, for example at ${\lambda _d} = 0.3$ ${N_{\max }} = 30$ for the vacuum and single-hole sectors of the Hilbert space and ${N_{\max }} = 50$ for the two-hole sector.~\cite{Makarov2015} Thus there are ${N_{\max }} + 1$ states of thermal phonons in the zero-hole sector, the oxygen atoms oscillate about their equilibrium position in the potential of crystal lattice.

The single-hole (${n_h} = 1$) cluster states can be written as
\begin{equation}
\left| {1\sigma ,i} \right\rangle  = \sum\limits_{\nu  = 0}^{N_{max}} {\left( {c_{i\nu }^d\left| {{d_\sigma }} \right\rangle \left| \nu  \right\rangle  + c_{i\nu}^b\left| {{b_\sigma }} \right\rangle \left| \nu  \right\rangle } \right)}
\label{es1}
\end{equation}
Here $\left| {{d_\sigma }} \right\rangle  = d_{{x^2} - {y^2}\sigma }^\dag \left| 0 \right\rangle $,  $\left| {{b_\sigma }} \right\rangle  = b_\sigma ^\dag \left| 0 \right\rangle $, index $i$ numerates the ground and excited single-hole eigenstates. Corresponding electronic configurations of stoichiometric La$_2$CuO$_4$ are $\left| {3{d^9}2{p^6}} \right\rangle$ and $\left| {3{d^{10}}2{p^5}} \right\rangle $.

Electron part of the two-hole (${n_h} = 2$) basis consists of the Zhang-Rice singlet state $\left| {{\rm{ZR}}} \right\rangle  = \left| {\frac{1}{{\sqrt 2 }}\left( {{d_ \downarrow }{b_ \uparrow } - {d_ \uparrow }{b_ \downarrow }} \right)} \right\rangle $, the two holes on copper ion state $\left| {{d_ \downarrow }{d_ \uparrow }} \right\rangle $ and the two holes on oxygen ion state $\left| {{b_ \downarrow }{b_ \uparrow }} \right\rangle $.The electronic configurations $\left| {3{d^9}2{p^5}} \right\rangle $, $\left| {3{d^8}2{p^6}} \right\rangle $, $\left| {3{d^{10}}2{p^4}} \right\rangle $ correspond to these states. Two-hole eigenstates of CuO$_6$ cluster with phonons are
\begin{eqnarray}
\left| {2,j} \right\rangle & = & \sum\limits_{\nu  = 0}^{N_{max}} \left(  c_{j\nu }^{ZR}\left| {{\rm{ZR}}} \right\rangle \left| \nu  \right\rangle + \right. \nonumber \\
 &+ & \left. c_{j\nu }^{dd}\left| {{d_{\downarrow }}{d_{\uparrow }}} \right\rangle \left| \nu  \right\rangle  + c_{j\nu }^{bb}\left| {{b_ \downarrow }{b_ \uparrow }} \right\rangle \left| \nu  \right\rangle \right)
\label{es2}
\end{eqnarray}
It is seen from Eq.~(\ref{es1}) and Eq.~(\ref{es2}) that each hole state of the pure electronic system transforms into ${N_{\max }} + 1$ vibronic states with taking into account phonon subsystem. A scheme of multielectron and multiphonon CuO$_6$ cluster levels (\ref{es0}), (\ref{es1}) and (\ref{es2}) is depicted in Fig.~\ref{fig:levels}. Without EPI all states are Cartesian production of hole and phonon wave functions and their energy levels are equidistant. With taking into account the EPI the equilibrium positions of oxygen atoms shift to new positions. Displacements of oxygen environment of copper atom are expressed by cloud of virtual phonons. Cluster eigenstates become polaron states i.e. they are superpositions of products of hole and phonon wave functions. Similar description of the local polaronic states has been discussed previously.~\cite{Piekarz1999} The number of the virtual phonon states in the cloud and its dependence on the EPI coupling has been discussed in detail for zero temperature in our paper Ref.~\onlinecite{Makarov2015}.

\begin{figure}
\center
\includegraphics[width=1.0\linewidth]{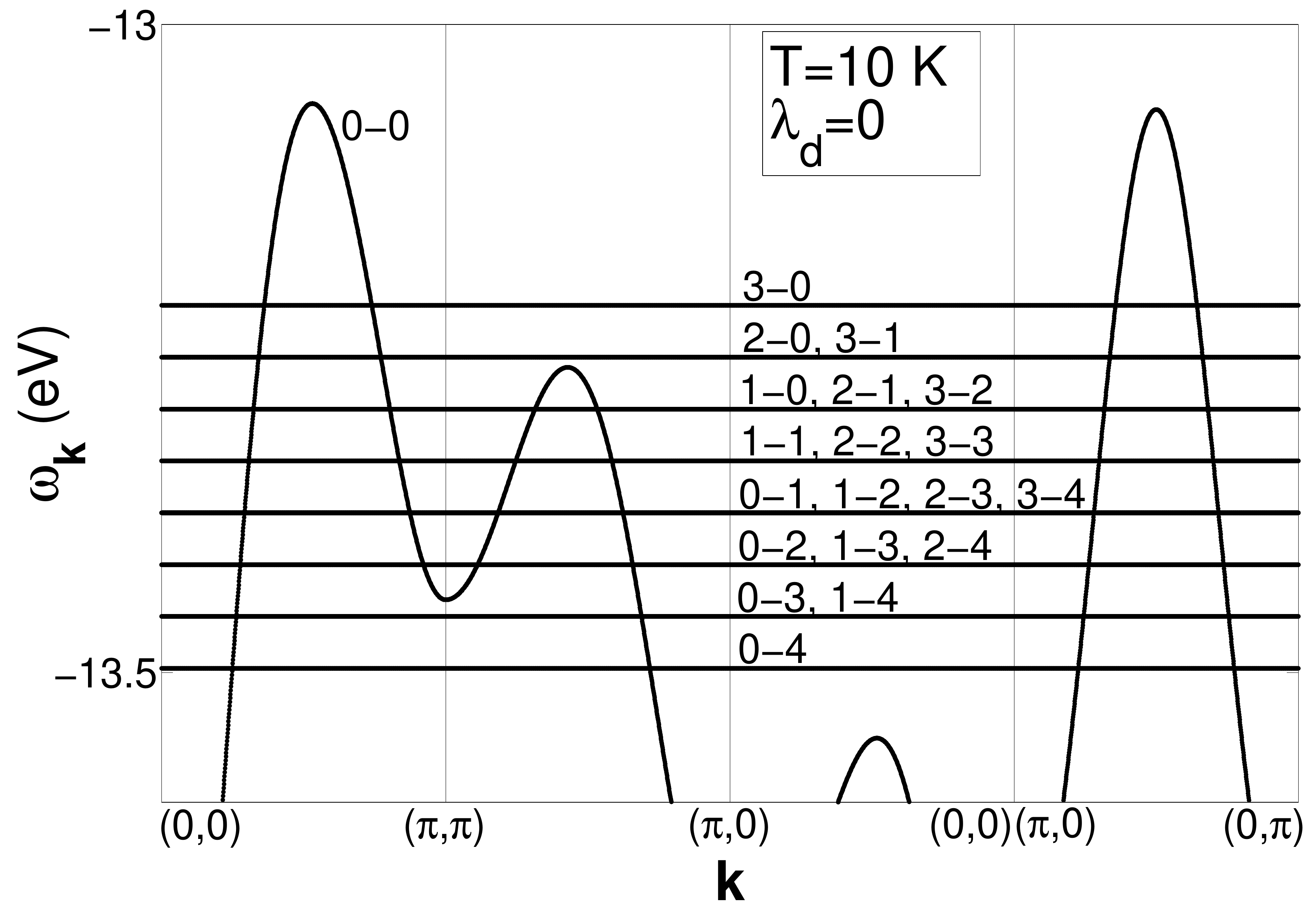}
\caption{\label{fig:FC_resonances} Dispersion of the upper part of the electron LHB without EPI at low temperature. In general the quasiparticle bands are characterized by spectral weight at each $k$-point but in this figure spectral weight difference is not shown. The quasparticle band $0-0$ (phononless excitation between ground single-hole and ground two-hole states) is dispersive due to the intercluster hopping. The Fermi-type multiphonon excitations (Franck-Condon resonances) are shown by a set of horizontal lines. They are dispersionless, multiply degenerate and has zero spectral weight at zero temperature and zero EPI constant. First (second) index in the notation of the quasiparticle denotes simultaneously index of single-hole (two-hole) cluster eigenstate.}
\end{figure}

The Fermi-type polaronic excitations (arrows in Fig.~\ref{fig:levels}) from cluster eigenstates $\left| {1,i} \right\rangle $ to $\left| {2,j} \right\rangle $ result in creation of hole, and from $\left| {1,i} \right\rangle $ to $\left| {0,\nu } \right\rangle $ results in creation of electron. Both change the number of fermions by unity, number of thermal phonons and polarization of oxygen environment of copper atom. These local polaron multiphonon excitations describe the Franck-Condon processes. Each excitation can be described by the Hubbard operator $X_{\bf{f}}^{pq} = \left| p \right\rangle \left\langle q \right|$, where $\left| q \right\rangle $ is the initial cluster eigenstate and $\left| p \right\rangle $ is the final cluster eigenstate. The Fermi-type operators of annihilation of hole on copper and oxygen orbital can be expressed in terms of the Fermi-type Hubbard operators $X_{\bf{f}}^{pq}$:
\begin{eqnarray}
{d_{{\bf{f}}\sigma }} = \sum\limits_{pq} {{\gamma _{d\sigma }}\left( {pq} \right)} X_{\bf{f}}^{pq} \nonumber \\
{b_{{\bf{f}}\sigma }} = \sum\limits_{pq} {{\gamma _{b\sigma }}\left( {pq} \right)} X_{\bf{f}}^{pq}
\label{Xop}
\end{eqnarray}
The phonon annihilation operator is expressed through the Bose-type Hubbard operators $Z_{\bf{f}}^{pp'}$:
\begin{equation}
{A_{\bf{f}}} = \sum\limits_{pp'} {{\gamma _A}\left( {pp'} \right)} Z_{\bf{f}}^{pp'}
\label{Zop}
\end{equation}
where states $\left| p \right\rangle $ and $\left| {p'} \right\rangle $ have the same number of holes and belong to the same sector of Hilbert space $\left| {{n_h}} \right\rangle $. Electronic excitations in the CuO$_6$ cluster are now described not by holes on $d$ or $b$ orbitals but polaronic quasiparticles with spectral weight defined by occupation of initial and final states of corresponding transition and their overlap. Since we will study the electronic states close to the top of the valence band that in our model is the lower Hubbard band (LHB), the Franck-Condon processes will be denoted by "index of single-hole eigenstate - index of two-hole eigenstate", $i - j$. Franck-Condon resonances are multiply degenerate, for instance single-phonon excitations $0-1$, $1-2$, $2-3$ etc have the same energy~(Fig.~\ref{fig:FC_resonances}).

Hamiltonians ${H_c}$ and ${H_{cc}}$ in Eq.(\ref{Hc_Hcc_tot}) are rewritten in the terms of Hubbard operators:
\begin{eqnarray}
& {H_c} & = \sum\limits_{\bf{f}} {\left[ {\sum\limits_l {{\varepsilon _{0l}}Z_{\bf{f}}^{0l,0l}}  + \sum\limits_i {{\varepsilon _{1i}}Z_{\bf{f}}^{1i,1i}}  + \sum\limits_j {{\varepsilon _{2j}}Z_{\bf{f}}^{2j,2j}} } \right]} \nonumber \\
& {H_{cc}} & = \sum\limits_{{\bf{f}} \ne {\bf{g}}} {\left[ {\sum\limits_{mn} {2{t_{pd}}{\mu _{{\bf{fg}}}}\gamma _{{d_x}}^ * \left( m \right){\gamma _b}\left( n \right)\mathop {X_{\bf{f}}^m}\limits^\dag  X_{\bf{g}}^n} } \right.}  - \nonumber \\
& &\left. { - \sum\limits_{mn} {2{t_{pp}}{\nu _{{\bf{fg}}}}\gamma _b^ * \left( m \right){\gamma _b}\left( n \right)\mathop {X_{\bf{f}}^m}\limits^\dag  X_{\bf{g}}^n} } \right]
\label{Hc_Hcc_Xop}
\end{eqnarray}
Here ${\varepsilon _{0l}}$, ${\varepsilon _{1i}}$, ${\varepsilon _{2j}}$ are the energies of cluster eigenstates with ${n_h} = 0,1,2$. The intercluster interactions result from the $p-d$ and $p-p$ hoppings of the Hubbard polarons between clusters, we consider hopping up to sixth neighbors. To obtain the band dispersion and spectral function of Hubbard polarons we use the equation of motion for the Green function ${D^{mn}}\left( {{\bf{f}},{\bf{g}}} \right) = \left\langle {\left\langle {{X_{\bf{f}}^m}}
 \mathrel{\left | {\vphantom {{X_{\bf{f}}^m} {X_{\bf{g}}^n}}}
 \right. \kern-\nulldelimiterspace}
\mathop {X_{\bf{g}}^n}\limits^\dag \right\rangle } \right\rangle $, where $m$,$n$ are the quasiparticle band indexes, this index is uniquely defined by initial and final states of excitation $m \equiv \left( {p,q} \right)$. The electron Green function ${G_{\lambda \lambda '}}\left( {{\bf{f}},{\bf{g}}} \right) = \left\langle {\left\langle {{{a_{\lambda {\bf{f}}}}}}
 \mathrel{\left | {\vphantom {{{a_{\lambda {\bf{}}f}}} {a_{\lambda '{\bf{g}}}^\dag }}}
 \right. \kern-\nulldelimiterspace}
 {{a_{\lambda '{\bf{g}}}^\dag }} \right\rangle } \right\rangle $ is connected with the quasiparticle Green function ${D^{mn}}\left( {{\bf{f}},{\bf{g}}} \right)$ by the following relation:
\begin{equation}
{G_{\lambda \lambda '}}\left( {{\bf{f}},{\bf{g}}} \right) = \sum\limits_{mn} {\gamma _{\lambda '}^ * \left( n \right){\gamma _\lambda }\left( m \right){D^{mn}}\left( {{\bf{f}},{\bf{g}}} \right)}
\label{GreenFunc_G_D}
\end{equation}
There are many quasiparticle excitations in our system therefore it is convenient to introduce matrix Green function $\hat D\left( {{\bf{f}},{\bf{g}}} \right)$ with matrix elements ${D^{mn}}\left( {{\bf{f}},{\bf{g}}} \right)$, indexes of row $m$ and column $n$ run over all quasiparticle bands. The set of equations of motion is decoupled in the generalized Hartri-Fock approximation by method of irreducible Green functions~\cite{Plakida1989,Yushankhay1991,Valkov2005,KorshOvch2007} taking into account the interatomic spin-spin correlation functions. The Dyson equation for the matrix Green function $\hat D\left( {{\bf{f}},{\bf{g}}} \right)$ in the momentum space has the form
\begin{equation}
\hat D\left( {{\bf{k}};\omega } \right) = {\left[ {\hat D_0^{ - 1}\left( \omega  \right) - \hat F\hat {\tilde t}\left( {{\bf{k}};\omega } \right) + \hat \Sigma \left( {{\bf{k}};\omega } \right)} \right]^{ - 1}}\hat F
\label{Dyson_eq}
\end{equation}
In this equation ${\hat D_0}$ is the exact local Green function, its matrix elements $D_0^{mn} = {{{\delta _{mn}}F\left( m \right)} \mathord{\left/
 {\vphantom {{{\delta _{mn}}F\left( m \right)} {\left( {\omega  - \Omega \left( m \right)} \right)}}} \right.
 \kern-\nulldelimiterspace} {\left( {\omega  - \Omega \left( m \right)} \right)}}$, $\Omega \left( m \right) = \Omega \left( {pq} \right) = {E_p} - {E_q}$ is the energy of the quasiparticle excitation between states $q$ and $p$, matrix elements $F^{mn} = F\left( m \right){\delta _{mn}}$, $F\left( m \right) = F\left( {pq} \right) = \left\langle {{X^{pp}}} \right\rangle  + \left\langle {{X^{qq}}} \right\rangle $ is the filling factor of the quasiparticle. The matrix of intercluster hopping $\hat {\tilde t}\left( {{\bf{k}};\omega } \right)$ is determined by $p-d$ and $p-p$ hoppings with the matrix elements $\tilde t_{\bf{k}}^{mn} = \sum\limits_{\lambda \lambda '} {\gamma _\lambda ^ * \left( m \right)} {\gamma _{\lambda '}}\left( n \right)\tilde t_{\bf{k}}^{\lambda \lambda '}$. $\hat \Sigma \left( {{\bf{k}};\omega } \right)$ is the self-energy operator which contains spin-spin correlation functions. The filling factor results in the temperature dependence of the quasiparticle band dispersion and the spectral weight.

\begin{figure}
\center
\includegraphics[width=1.0\linewidth]{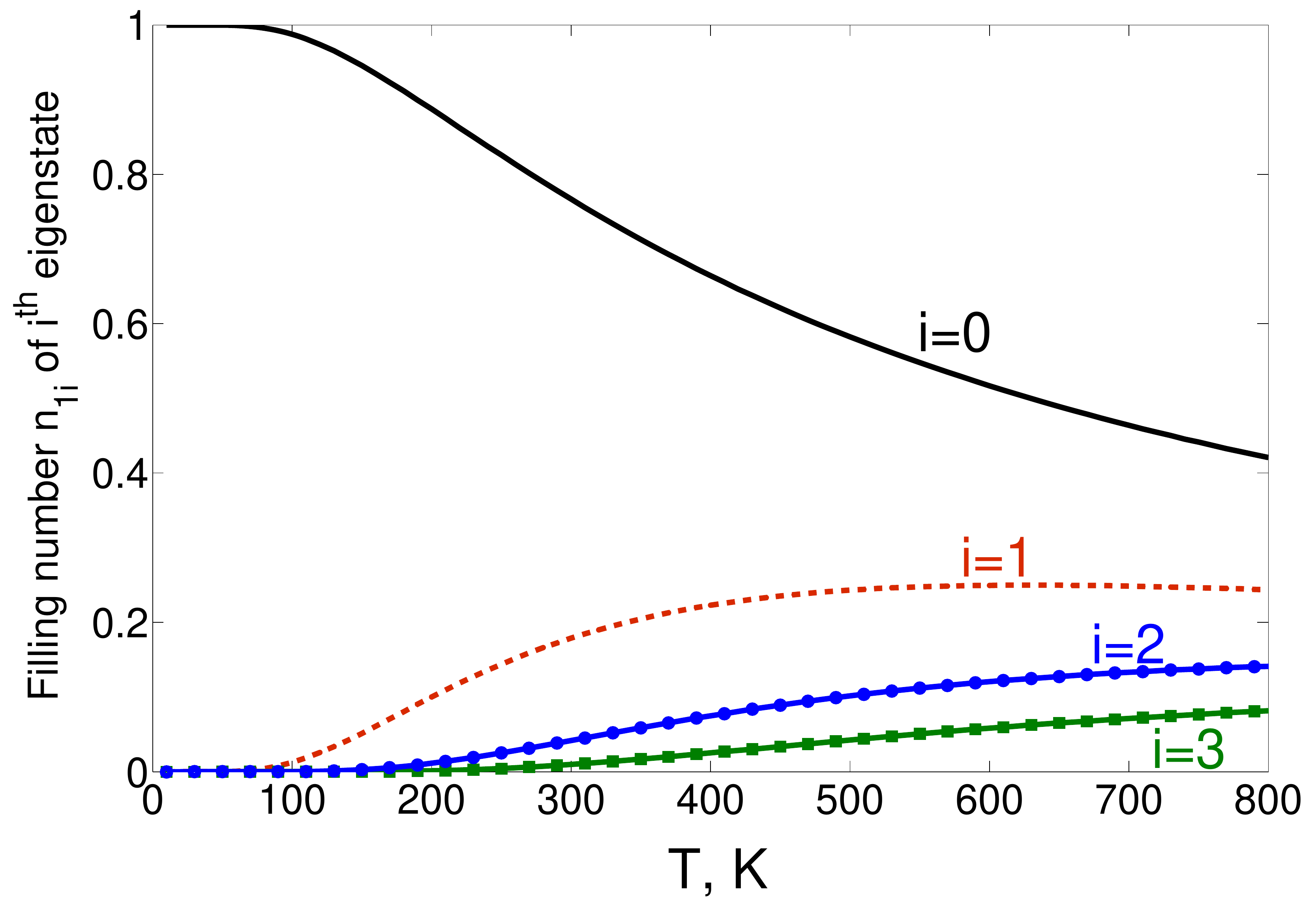}
\caption{\label{fig:nzap_Tdep} Temperature dependence of the filling numbers ${n_{1i}}$ of single-hole eigenstates $\left| {1,i} \right\rangle $ of the CuO$_6$ cluster. It is seen that at $T = 0$ K only the ground state $\left| {1,0} \right\rangle $ is occupied. Hole distributes over a number of excited states with increasing temperature.}
\end{figure}

Filling of cluster eigenstates is determined self-consistently from the condition of completeness $\sum\limits_p {X_{\bf{f}}^{pp} = 1} $ and the chemical potential equation (here $n = 1 + x$ is the hole concentration for La$_{2-x}$Sr$_x$CuO$_4$):
\begin{eqnarray}
n &=& 1 + x = \nonumber \\
&=& \sum\limits_l {0 \cdot \left\langle {{X^{0l,0l}}} \right\rangle }  + \sum\limits_i {1 \cdot \left\langle {{X^{1i,1i}}} \right\rangle }  + \sum\limits_j {2 \cdot \left\langle {{X^{2j,2j}}} \right\rangle } \nonumber \\
\label{chempot_eq}
\end{eqnarray}
where $l,i,j$ runs over all zero-, single- and two-hole eigenstates respectively, and from the
Boltzmann distribution for a given temperature:
\begin{equation}
{n_{1i}} = {n_{10}}\exp \left( { - \frac{{{E_{1i}} - {E_{10}}}}{{kT}}} \right)
\label{Gibbs}
\end{equation}
Below we will calculate the quasiparticle band structure with the following parameters of Hamiltonian (in eV):
\begin{eqnarray}
{\varepsilon _d} = 0, {\varepsilon _p} = 1.5, {t_{pd}} = 1.36, {t_{pp}} = 0.86 \nonumber \\
{U_d} = 20, {U_p} = 18, {V_{pd}} = 14
\label{parameters}
\end{eqnarray}
Here the tight binding parameters (intraatomic energies and nearest neighbor hopping) for La$_2$CuO$_4$ were obtained from LDA+GTB method.~\cite{LDA+GTB} As concerns the Coulomb parameters we will study two different regimes, the over correlated one with all Coulomb parameters increased more than twice (\ref{parameters}). Below in Section~\ref{T_dep_with_EPI_real_par} we present the polaronic band structure with realistic Coulomb parameters (\ref{realistic_parameters}), obtained from the constrained-LDA calculations.~\cite{LDA+GTB}

At $T=0$ K only the ground single-hole state is filled (Fig.~\ref{fig:nzap_Tdep}) and hence only excitations involving the single-hole ground state $\left| {1,0} \right\rangle $ have a non-zero spectral weight (solid arrows in Fig.~\ref{fig:levels}). Filling of ground state falls whereas occupation of $1^{th}$, $2^{th}$ and $3^{rd}$ excited single-hole states monotonically grows with increasing temperature (Fig.~\ref{fig:nzap_Tdep}) according to Eq.~(\ref{Gibbs}). At $T=800$ K occupation is significant up to $7^{th}$ excited state, ${n_4} = 0.02$, ${n_5} = 0.01$, ${n_6} = 0.007$, ${n_7} = 0.003$. Transitions involving occupied single-hole excited states acquire spectral weight (dashed arrows in Fig.~\ref{fig:levels}).

It is known that there is the long-range antiferromagnetic order in the undoped cuprates at low temperatures. In GTB approach the temperature dependent sublattice magnetization determines the value of spin-up/spin-down splitting resulting in the temperature dependent electronic structure.~\cite{MakOvchJETP2015} Here we will investigate electronic structure in wide interval of temperatures including the paramagnetic phase above the Neel temperature. The short-range antiferromagnetic order above the Neel temperature is still strong in the quasi-two-dimensional magnetic subsystem of La$_2$CuO$_4$. We will describe the short-range order in the paramagnetic state as the isotropic spin liquid with zero spin projections and non-zero spin-spin correlations functions following to Refs.~\onlinecite{Shimahara1991,Barabanov1994,Valkov2005}. The comparison of electronic structure calculated for the spin liquid state with spin-spin correlation functions $\left\langle {X_f^{\sigma \bar \sigma }X_g^{\bar \sigma \sigma }} \right\rangle $ for very low doping $x=0.01$ (Ref.~\onlinecite{KorshOvch2007}, FIG.3) and band structure obtained for undoped system with long-range order~\cite{MakOvchJETP2015} shows almost identical pictures. This similarity results from the large spin correlation length of the short-ordered antiferromagnetic clusters that exist at low doping. The dispersion of a hole hopping at the long-range antiferromagnetic background is very similar to the dispersion of a hole hopping at the local antiferromagnetic background with spin correlation length $\xi  \gg a$, $a$ is the lattice parameter. Therefore we will use short-range magnetic order given by spin-spin correlation functions and each single-hole level has spin degeneracy. Spin-spin correlation functions for doping $x=0.01$ at $T=0$ K was taken from Ref.~\onlinecite{KorshOvch2007} and their temperature dependence is taken from (Ref.~\onlinecite{Shimahara1991}). Spin-spin correlation functions decrease with temperature increasing. Significant damping of spin correlations takes place at $T \sim J = {{2{{\tilde t}^2}} \mathord{\left/
 {\vphantom {{2{{\tilde t}^2}} {{E_{gap}}}}} \right.
 \kern-\nulldelimiterspace} {{E_{gap}}}}$. At $T \sim J$ spin-spin correlator for nearest and next nearest neighbors is $60$ percent and $30$ percent of the value at zero temperature respectively.

\section{Temperature dependence of the electronic structure in the system without EPI \label{T_dep_without_EPI}}
\begin{figure*}
\center
\includegraphics[width=0.45\linewidth]{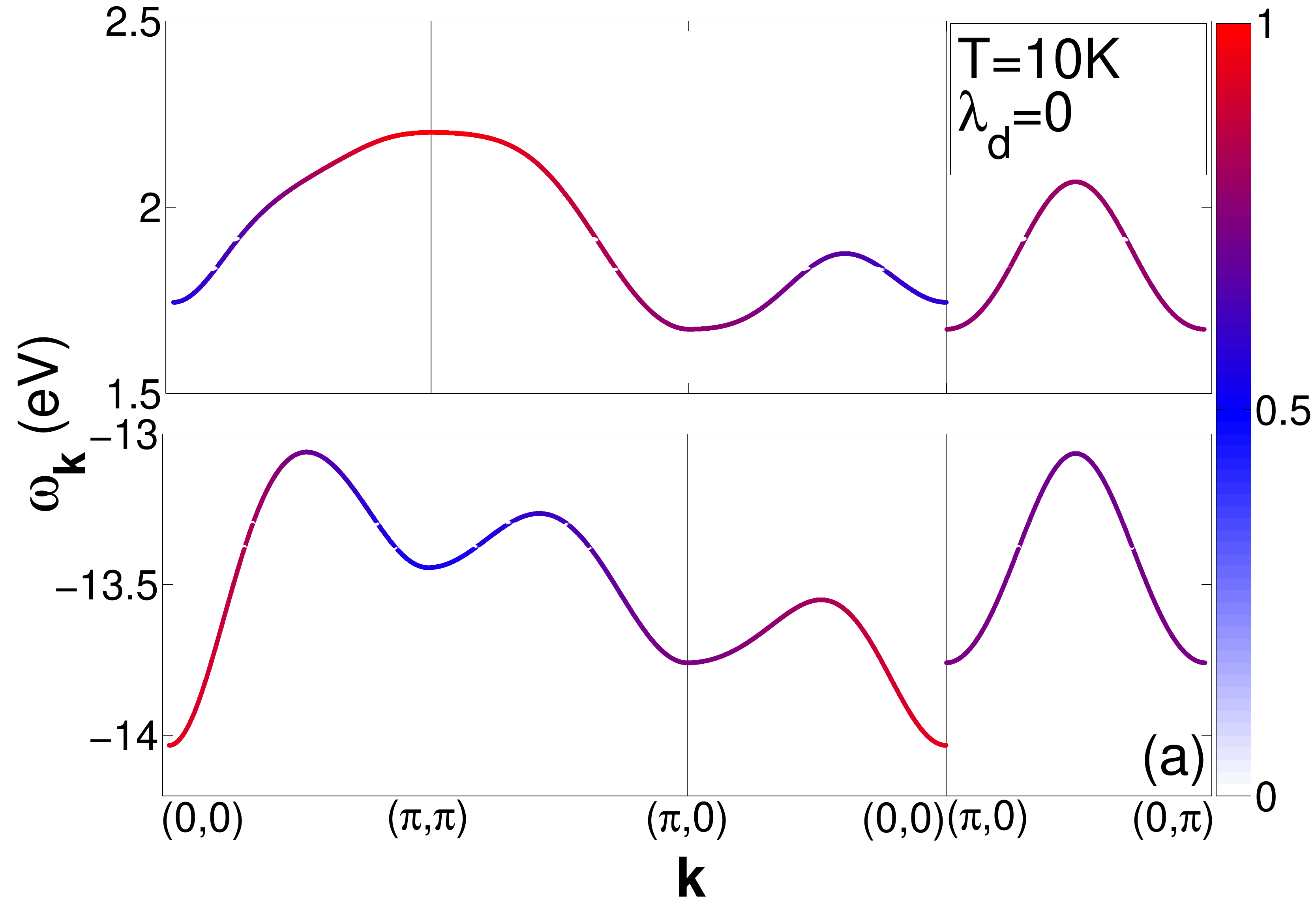}
\includegraphics[width=0.45\linewidth]{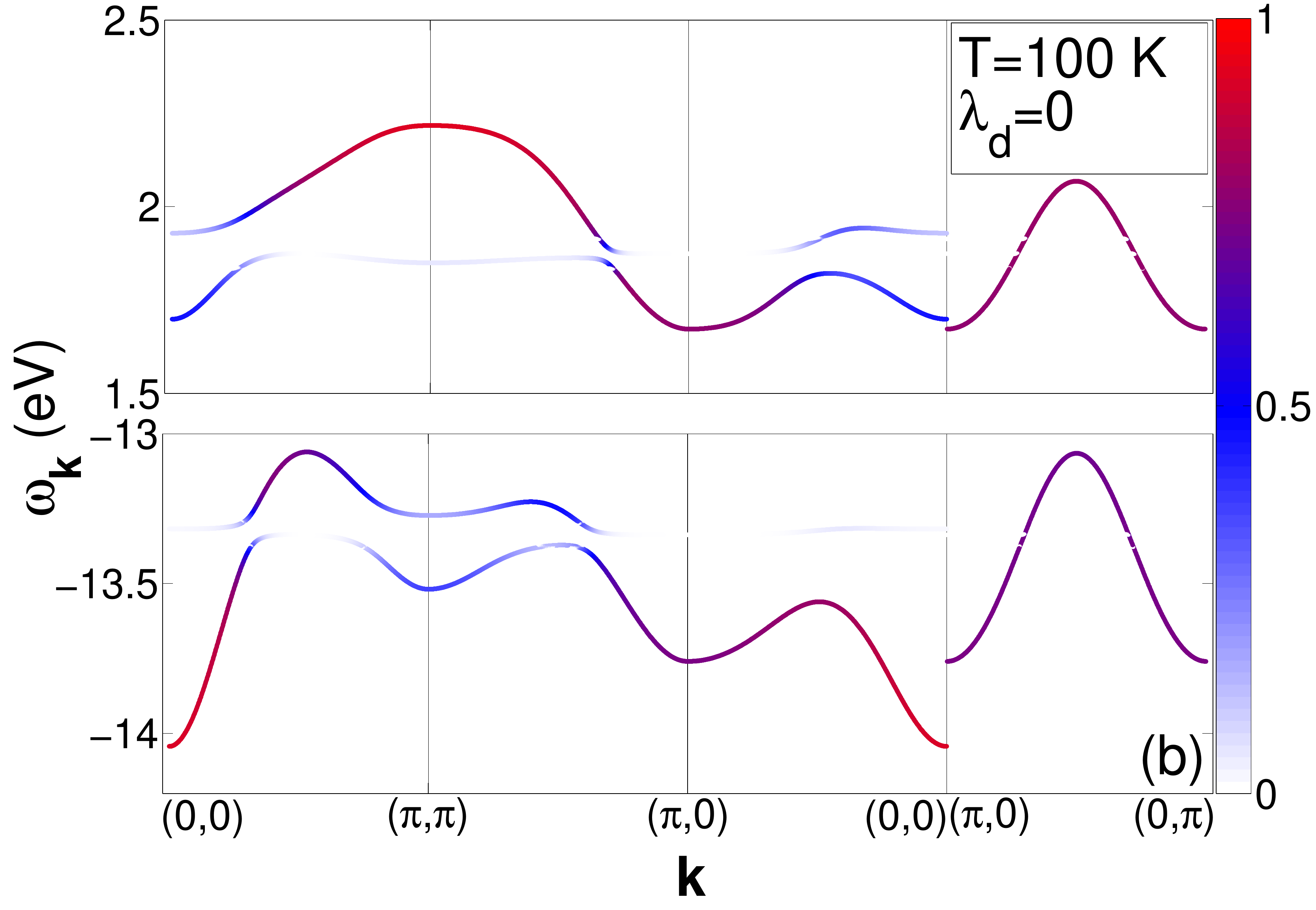}
\includegraphics[width=0.45\linewidth]{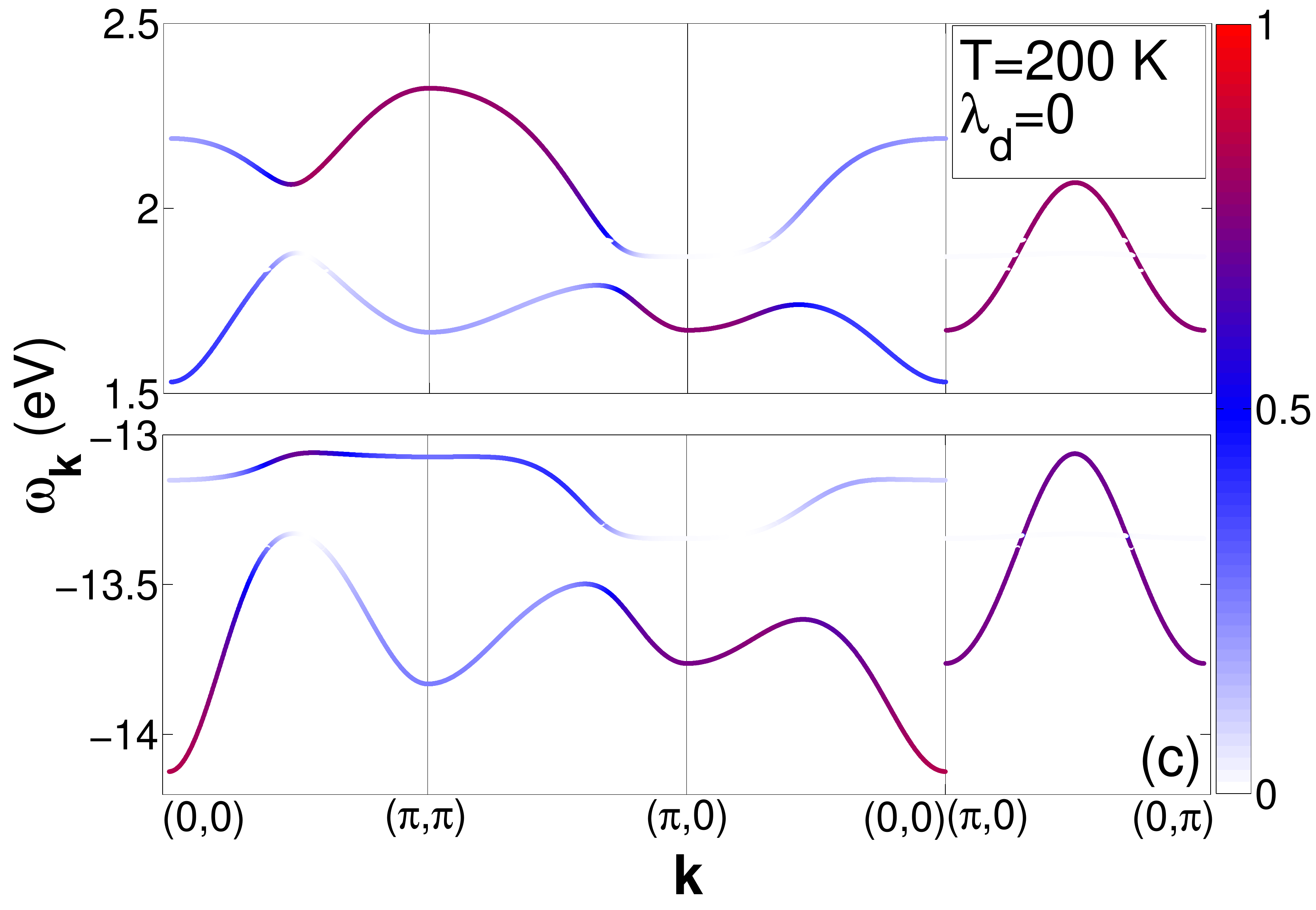}
\includegraphics[width=0.45\linewidth]{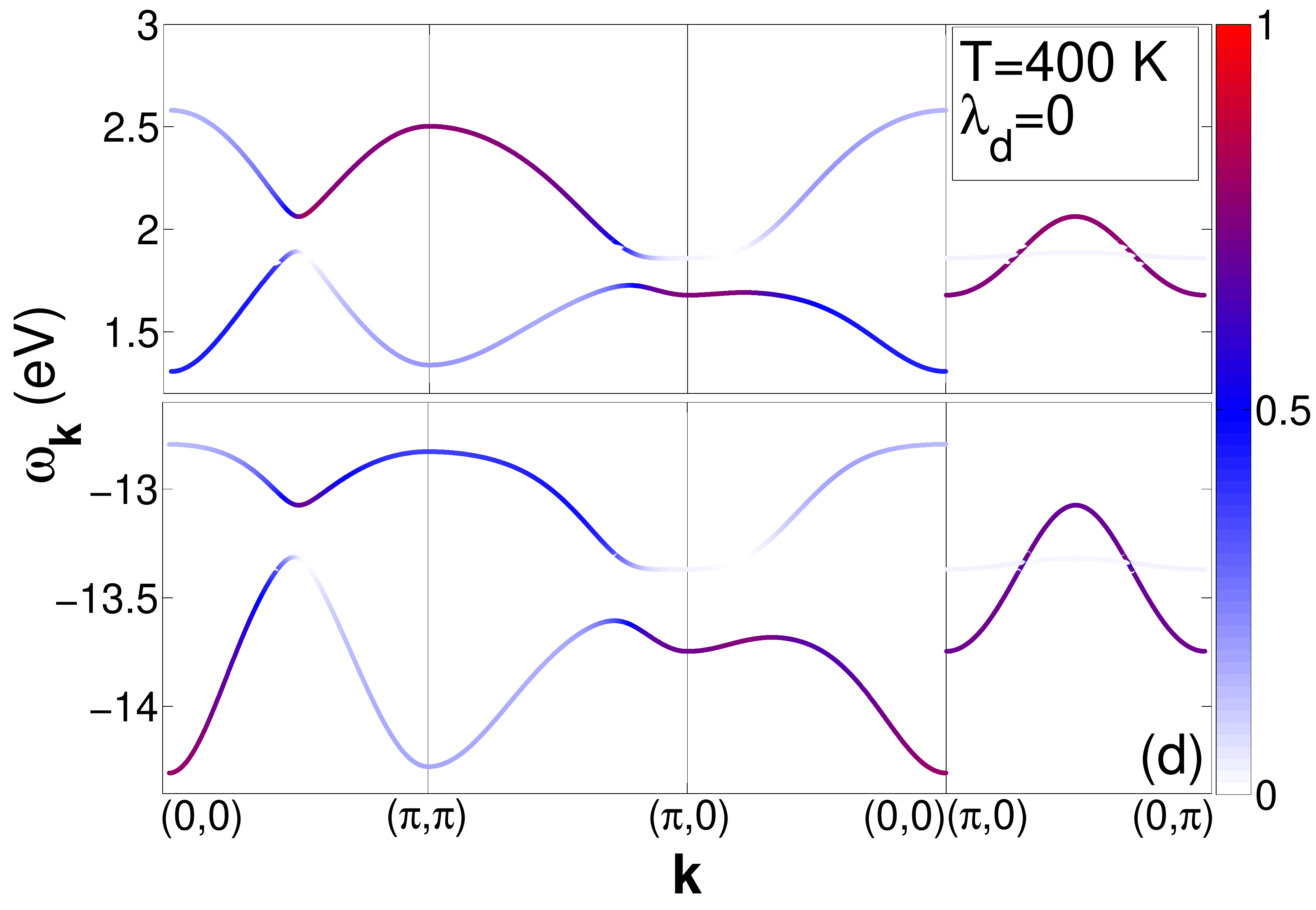}
\caption{\label{fig:bandstr_Tdep_withoutEPI} Evolution of the band structure of the quasiparticle excitations with increasing temperature for the system without EPI (${\lambda _d} = 0$). (a) $T = 10$ K, (b) $T = 100$ K, (c) $T = 200$ K, (d) $T = 400$ K. The conductivity (valence) bands are shown on the upper (lower) panel of each figure. Color in each $k$-point indicates the spectral weight of quasiparticle.}
\end{figure*}

Without EPI the LHB band structure is formed by $0-0$ quasiparticle (excitation between single-hole and two-hole ground states without phonons) band. There is also a set of dispersionless degenerate Franck-Condon resonances that at $T=0$ without EPI have zero spectral weight. Absence of spectral weight of these Franck-Condon resonances results from the orthogonality of phonon wave functions of the initial and final states. Energies of these resonances are determined relative to $\Omega \left( {0 - 0} \right)$ and are shown at Fig.~\ref{fig:FC_resonances} for clarity. In this Section we discuss a very specific effect of finite temperature on the electronic structure and spectral weight of polarons even without EPI. The matrix elements of the Fermi-type excitation between initial single-hole and final two-hole states with the same number of phonons (Franck-Condon processes $1-1$, $2-2$, $3-3$ etc) are equal to the same matrix element for $0-0$ process. The thermal occupation of the excited single-hole states shown in Fig.~\ref{fig:nzap_Tdep} and non-zero matrix elements due to the interband hopping ${\tilde t^{mn}}\left( {m \ne n} \right)$ will result in the hybridization of the Hubbard band $0-0$ and the phononless Franck-Condon excitations $1-1$, $2-2$, $3-3$ etc (Fig.~\ref{fig:bandstr_Tdep_withoutEPI}). These excitations are degenerate in energy, their energies $\Omega \left( {1 - 1} \right) = \Omega \left( {2 - 2} \right) = ... = \Omega \left( {{n_{1i}} - {n_{2i}}} \right) = ... = \Omega \left( {{N_{max}} - {N_{max}}} \right)$. Spectral weight of these pure electronic phononless excitations is zero at $T=0$ and grows with increasing temperature. Spectral weight of electron is redistributed among different excitations $0-0$, $1-1$, $2-2$ etc. As a result of hybridization of quasiparticles the $0-0$ band is splitted at the energy $\Omega \left( {1 - 1} \right)$ on two parts (Fig.~\ref{fig:bandstr_Tdep_withoutEPI}(b)). Splitting on two bands doesn't occur over whole band, it is absent in the direction $\left( {\pi ,0} \right) - \left( {0,\pi } \right)$. The higher the temperature the larger number of excitations split $0-0$ band and thus reconstruction of the band becomes more complicated. The upper part of the valence band is strongly modified under temperature increasing. Near $T=200$ K the local maximum at the point ${\bf{k}} = \left( {{\pi  \mathord{\left/
 {\vphantom {\pi  2}} \right.
 \kern-\nulldelimiterspace} 2},{\pi  \mathord{\left/
 {\vphantom {\pi  2}} \right.
 \kern-\nulldelimiterspace} 2}} \right)$ becomes local minimum (Fig.~\ref{fig:bandstr_Tdep_withoutEPI}(c)) by energy increase at points $\Gamma  = \left( {0,0} \right)$ and ${\rm M} = \left( {\pi ,\pi } \right)$. Further temperature increase results in energy growth at $\Gamma $ and ${\rm M}$ points (Fig.~\ref{fig:bandstr_Tdep_withoutEPI}(d)). The width of the whole band is increased.

\section{Temperature dependence of the electronic structure of the polaron quasiparticles \label{T_dep_with_EPI}}
\begin{figure*}
\center
\includegraphics[width=0.45\linewidth]{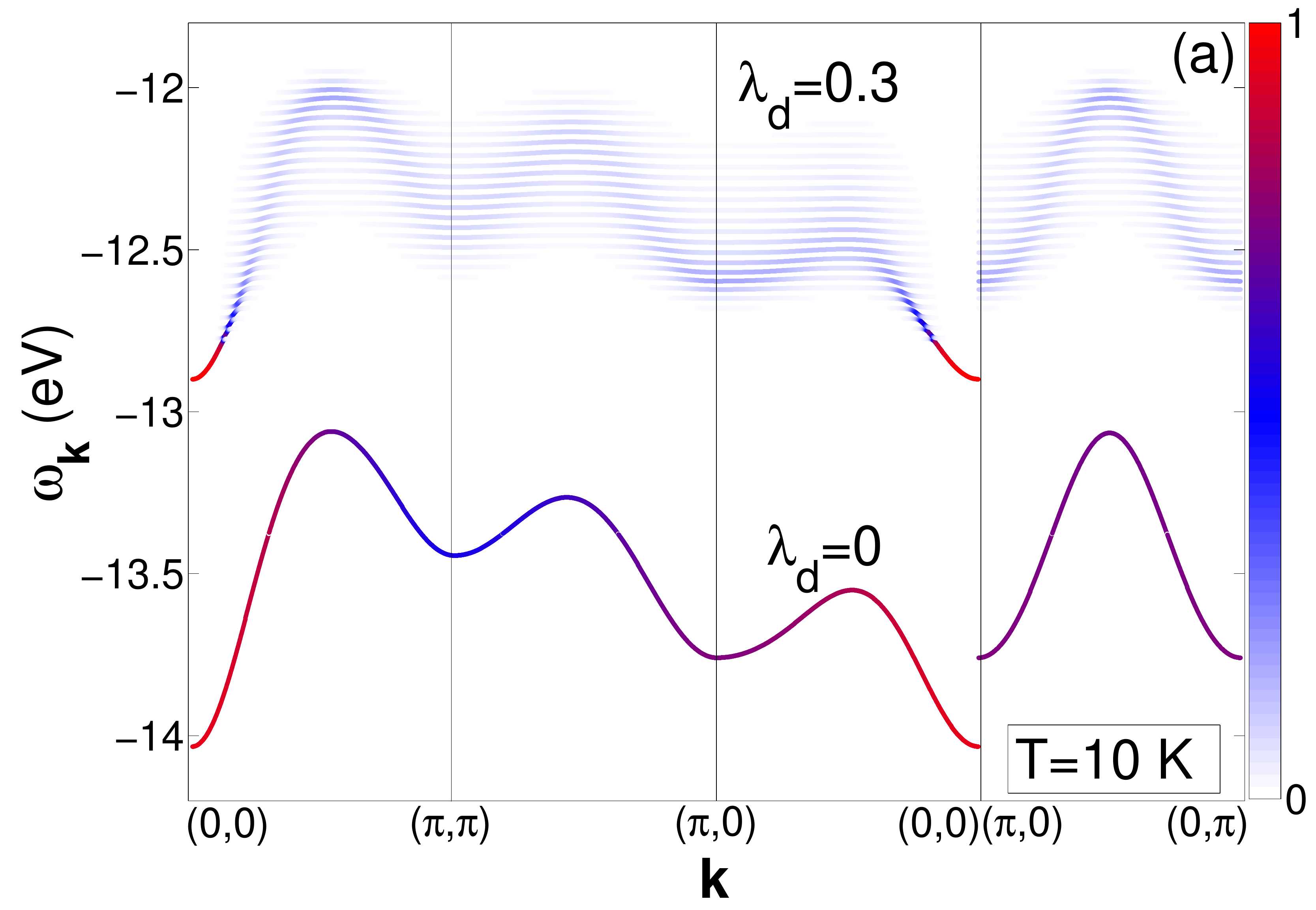}
\includegraphics[width=0.45\linewidth]{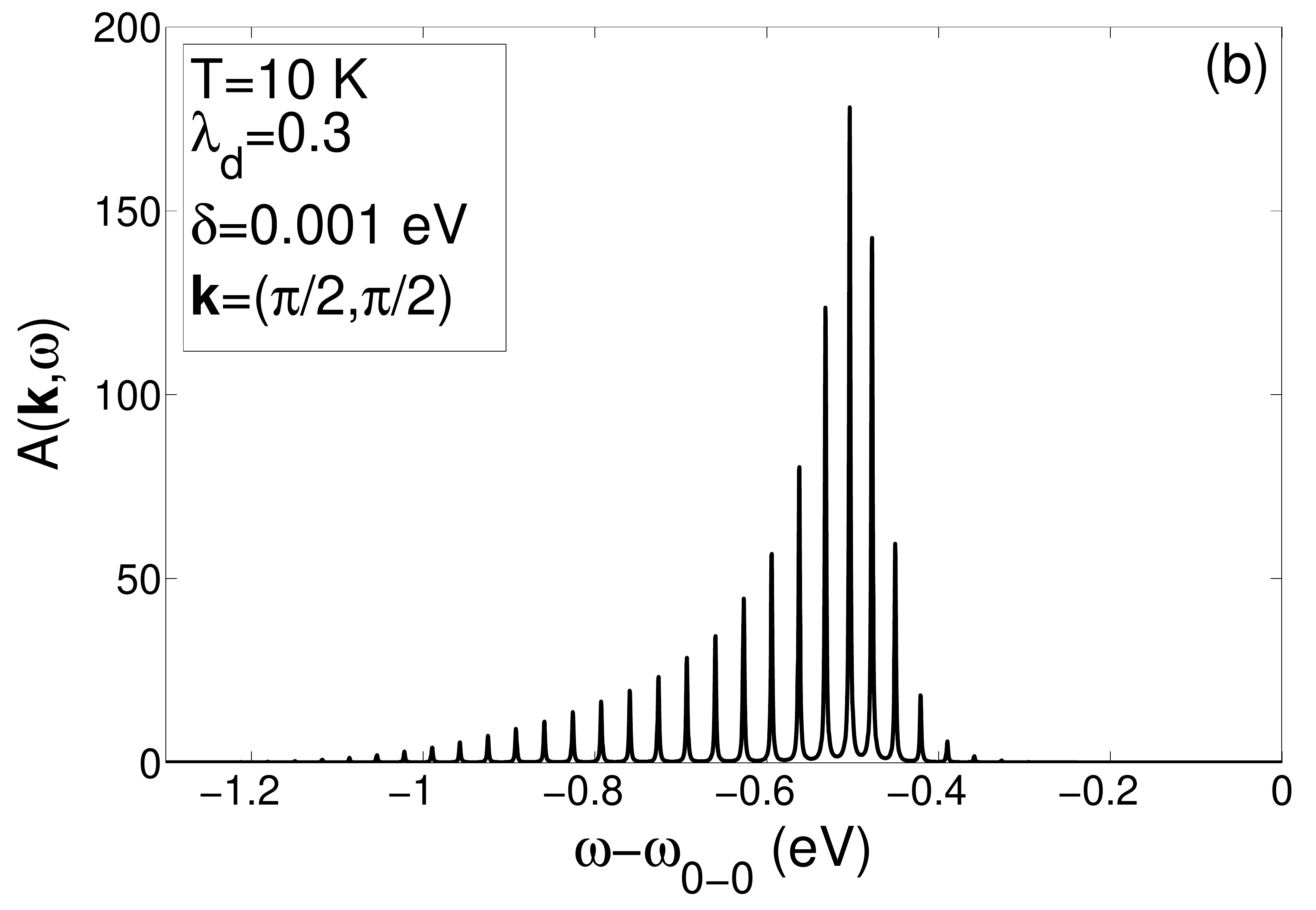}
\caption{\label{fig:bandstr_withEPI} Electronic structure of the LHB at low temperature ($T = 10$ K). (a) Band structures of the LHB at ${\lambda _d} = 0$ (energy region from $-14$ eV to $-13$ eV) and ${\lambda _d} = 0.3$ (energy region from $-13$ to $-12$ eV). Band structure of the LHB at ${\lambda _d} = 0.3$ consists of a number of Hubbard polaron subbands with weak dispersion. Color in each point of band structure is proportional to quasiparticle spectral weight. (b) Electron spectral function at ${\lambda _d} = 0.3$ and at ${\bf{k}} = \left( {\frac{\pi }{2},\frac{\pi }{2}} \right)$. It is formed by a number of peaks of the multiphonon Franck-Condon excitations. Electron energy is shifted from the Franck-Condon resonance $0-0$ that is the coherent band in the absence of the EPI. At large EPI the $0-0$ quasiparticle has zero spectral weight.}
\end{figure*}

\begin{figure*}
\center
\includegraphics[width=0.45\linewidth]{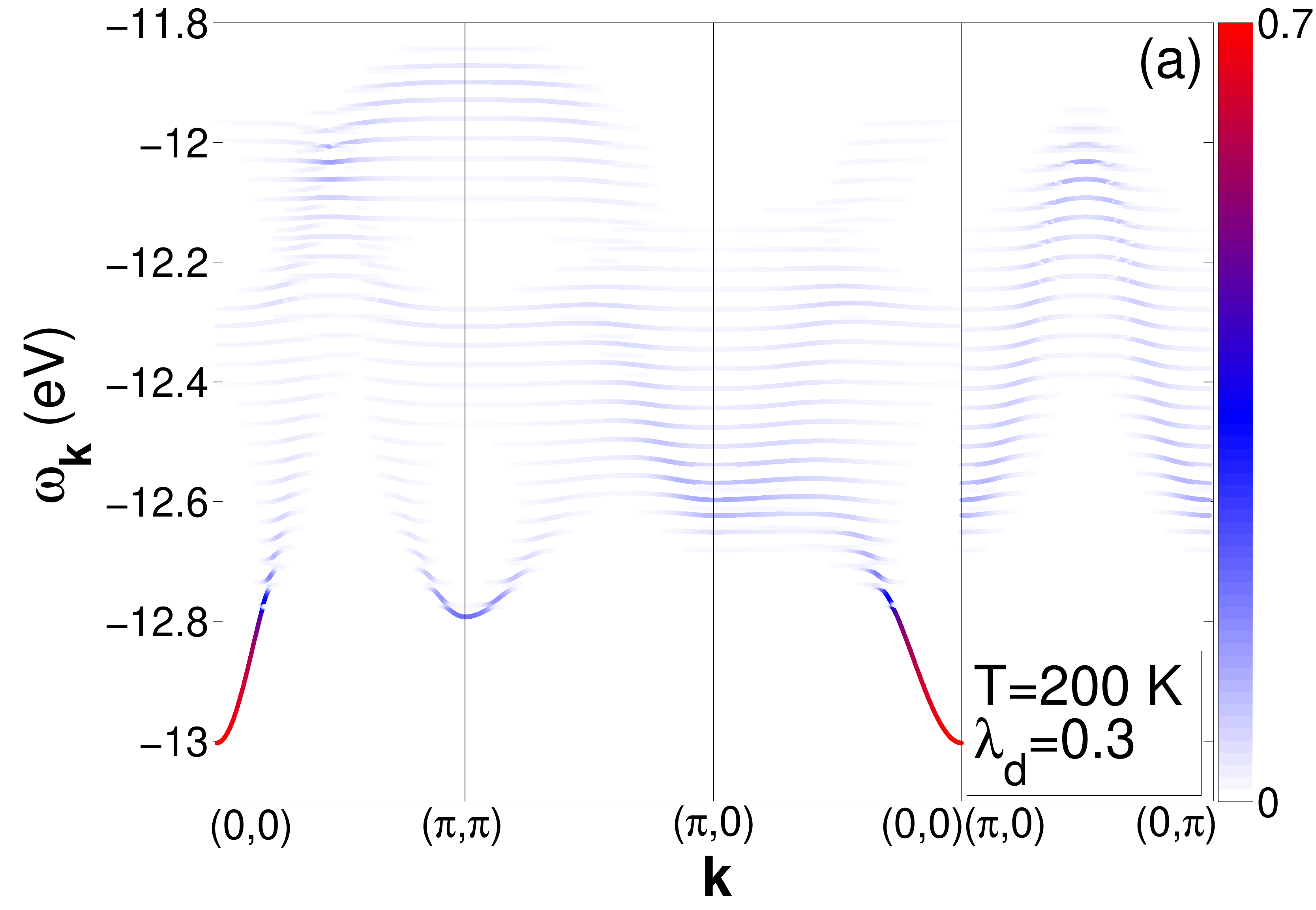}
\includegraphics[width=0.45\linewidth]{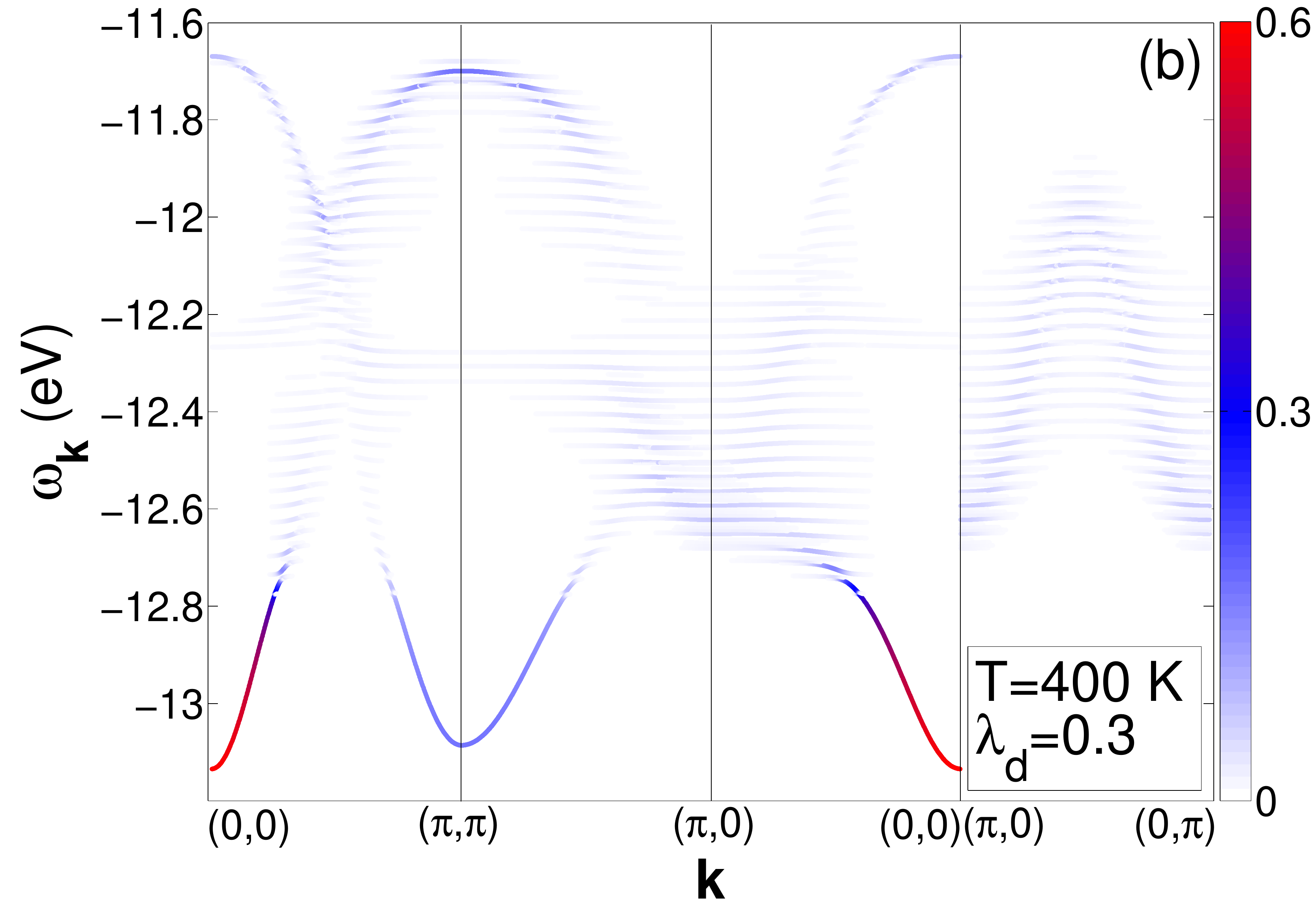}
\includegraphics[width=0.45\linewidth]{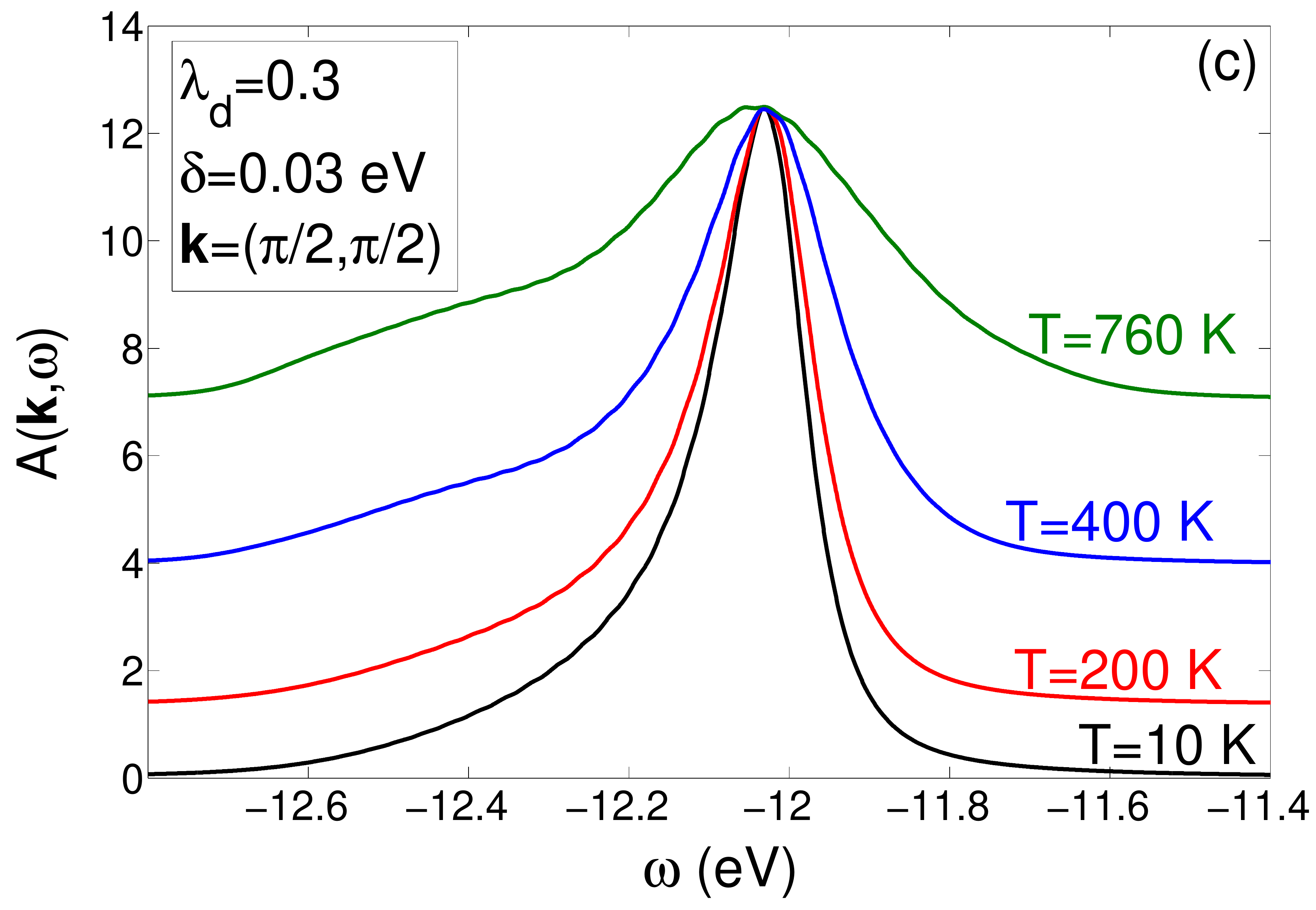}
\includegraphics[width=0.45\linewidth]{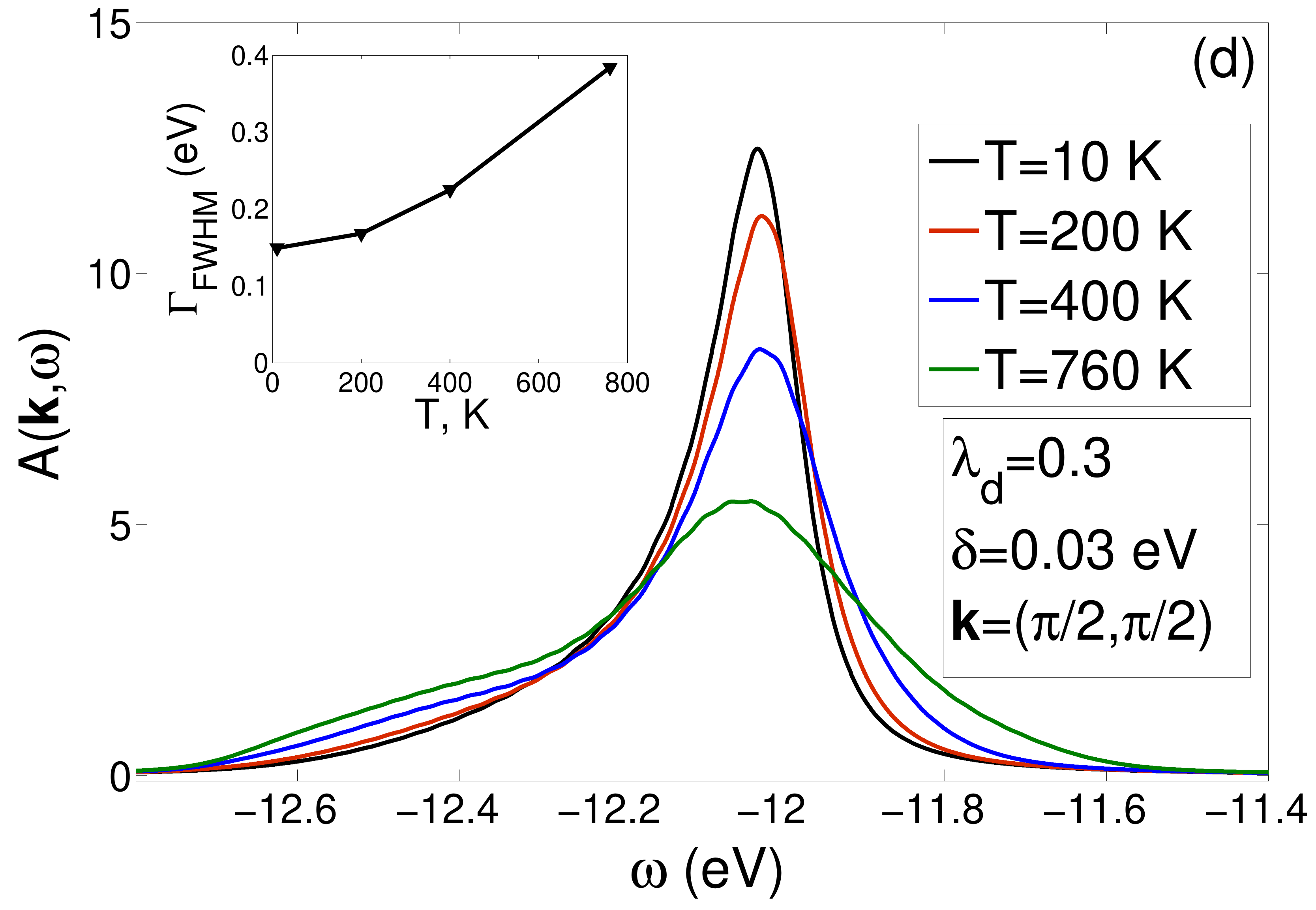}
\caption{\label{fig:elstr_Tdep_withEPI} Evolution of electronic structure at EPI constant ${\lambda _d} = 0.3$ with increasing temperature. Band structure of the valence band with spectral weight of quasiparticles in each $k$-point (spectral weight is displayed by color) at (a) $T = 200$ K, (b) $T = 400$ K. Note that spectral weight scale decreases with increasing temperature. (c),(d) The spectral functions and its FWHM (inset of (d)) at ${\bf{k}} = \left( {\frac{\pi }{2},\frac{\pi }{2}} \right)$ for different temperatures ($T = 10$ K, $T = 200$ K, $T = 400$ K, $T = 760$ K). (c) Broadening of spectral function peak with increasing temperature is shown, all curves are rescaled to have the same maximal value. (d) Damping of the spectral function peak intensity and shift of the peak are presented.}
\end{figure*}

\begin{figure}
\center
\includegraphics[width=1.0\linewidth]{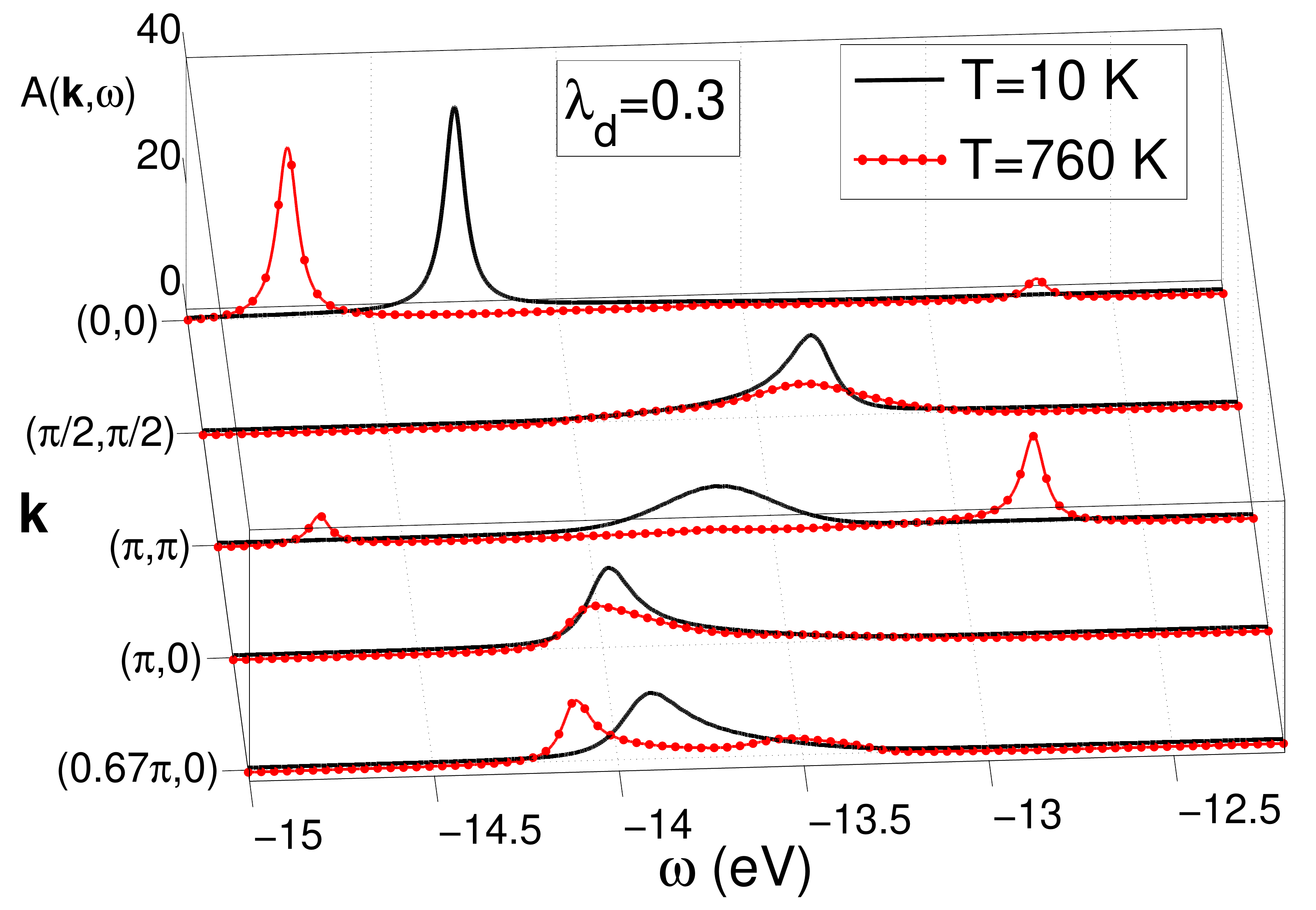}
\caption{\label{fig:shift_Tdep_withEPI} Polaron spectral functions at $T = 10$ K and $T = 760$ K for ${\lambda _d} = 0.3$. The temperature increasing results in small intensity satellites and the shift of the main peak. Spectral function is calculated with Lorentzian width $\delta  = 0.03$ eV.}
\end{figure}

The EPI results in the hybridization of the dispersive bands of the Hubbard electrons with the local Franck-Condon resonances forming the band structure of the Hubbard polarons. Probability of multiphonon excitations such as $0-1$, $0-2$, $0-3$ etc. becomes nonzero. These excitations acquire spectral weight, weak dispersion and begin to interact with $0-0$ polaron. Therefore a number of Hubbard polaron subbands in the valence band (and in the conductivity band) appears (Fig.~\ref{fig:bandstr_withEPI}(a)). Electron spectral weight is distributed over all possible polaron subbands decreasing the average spectral weight per one subband. It transfers from the coherent dispersive high-intensity $0-0$ quasiparticle to the incoherent multiphonon excitations with the EPI coupling increasing. Inside each Hubbard polaron subband the spectral weight is inhomogeneously distributed among all points in the momentum space. Electron spectral function
\begin{eqnarray}
A\left( {{\bf{k}},\omega } \right) & = &\left( { - \frac{1}{\pi }} \right) \sum \limits_{\lambda \lambda '\sigma mn} {\gamma _{\lambda '\sigma }^ * \left( n \right){\gamma _{\lambda \sigma }}\left( m \right) \times } \nonumber \\
& \times & {\mathop{\rm Im}\nolimits} {\left\langle {\left\langle {{X_{\bf{k}}^m}}
 \mathrel{\left | {\vphantom {{X_{\bf{k}}^m} {\mathop {X_{\bf{k}}^n}\limits^\dag  }}}
 \right. \kern-\nulldelimiterspace}
 {{\mathop {X_{\bf{k}}^n}\limits^\dag  }} \right\rangle } \right\rangle _{\omega  + i\delta }}
\label{SpectralFunction}
\end{eqnarray}
at each $k$-point is formed by several peaks reflecting the Franck-Condon excitations (Fig.~\ref{fig:bandstr_withEPI}(b)). Taking into account finite lifetime of quasiparticles the multipeak structure of spectral function can merge into single broad peak if the EPI constant is large enough. Form and width of this peak depends on the EPI value. In spite of large number of the Franck-Condon excitations crossing and lost of the coherency the LHB original dispersion of electronic model without EPI is mainly preserved, as can be seen by comparison the coherent LHB at ${\lambda _d} = 0$ and incoherent polaron band at ${\lambda _d} = 0.3$ in Fig.~\ref{fig:bandstr_withEPI}(a). This comparison also has revealed quite large shift of the polaron dispersion as a whole.

At zero temperature the main peak of spectral function is formed by excitation from the occupied ground single-hole eigenstate to excited multiphonon polaron two-hole state for which the Franck-Condon factor is largest. Filling of excited single-hole states with one, two, three etc. thermal phonons grows with increasing temperature (Fig.~\ref{fig:nzap_Tdep}) and such excitations as $1-0$, $2-0$, $1-2$ etc acquire spectral weight. Spectral weight of the quasiparticles which are primary formed by these new multiphonon excitations involving excited eigenstates also increases. Generally contributions of different multiphonon excitations to complex quasiparticle forming specific polaron subband are redistributed with changing temperature. Peaks of emerged quasiparticles are satellites of the main peak. Intensity of the main peak falls, intensity of satellites increases with temperature. Redistribution of spectral weight between quasiparticles involving ground and quasiparticles involving excited single-hole eigenstates causes the changes in width and shape of the resulting spectral function peak. Splitting of the LHB on the number of Hubbard polaron subbands and redistribution of spectral weight over these subbands occurs in addition to temperature transformation of $0-0$ quasiparticle band (Fig.~\ref{fig:elstr_Tdep_withEPI}(a),(b)) that has been discussed in the previous Section~\ref{T_dep_without_EPI}. For large EPI ${\lambda _d} = 0.3$ all spectral weight results from the multiphonon excitations. The main effects of temperature growth are broadening of spectral function (with Lorentzian width $\delta  = 0.03$ eV) (Fig.~\ref{fig:elstr_Tdep_withEPI}(c)) and damping of its maximal intensity (Fig.~\ref{fig:elstr_Tdep_withEPI}(d)). Value of the FWHM ${\Gamma _{FWHM}}$ grows in $1.3$ times with increasing temperature from $T = 200$ K to $T = 400$ K (inset if Fig.~\ref{fig:elstr_Tdep_withEPI}(d)) whereas in the ARPES ${\Gamma _{FWHM}}$ is doubled.~\cite{ShenKM2007} For the temperatures from $T = 200$ K to $T = 760$ K our calculations show that ${\Gamma _{FWHM}}$ increases more than twice. Thus broadening of spectral function with increasing temperature is qualitatively in agreement to ARPES spectra. Stronger influence of temperature on redistribution of filling numbers of local polaron multiphonon states is necessary to provide better agreement between calculations and ARPES data since the main mechanism of broadening related to redistribution of cluster eigenstates occupation.

\begin{figure*}
\center
\includegraphics[width=0.45\linewidth]{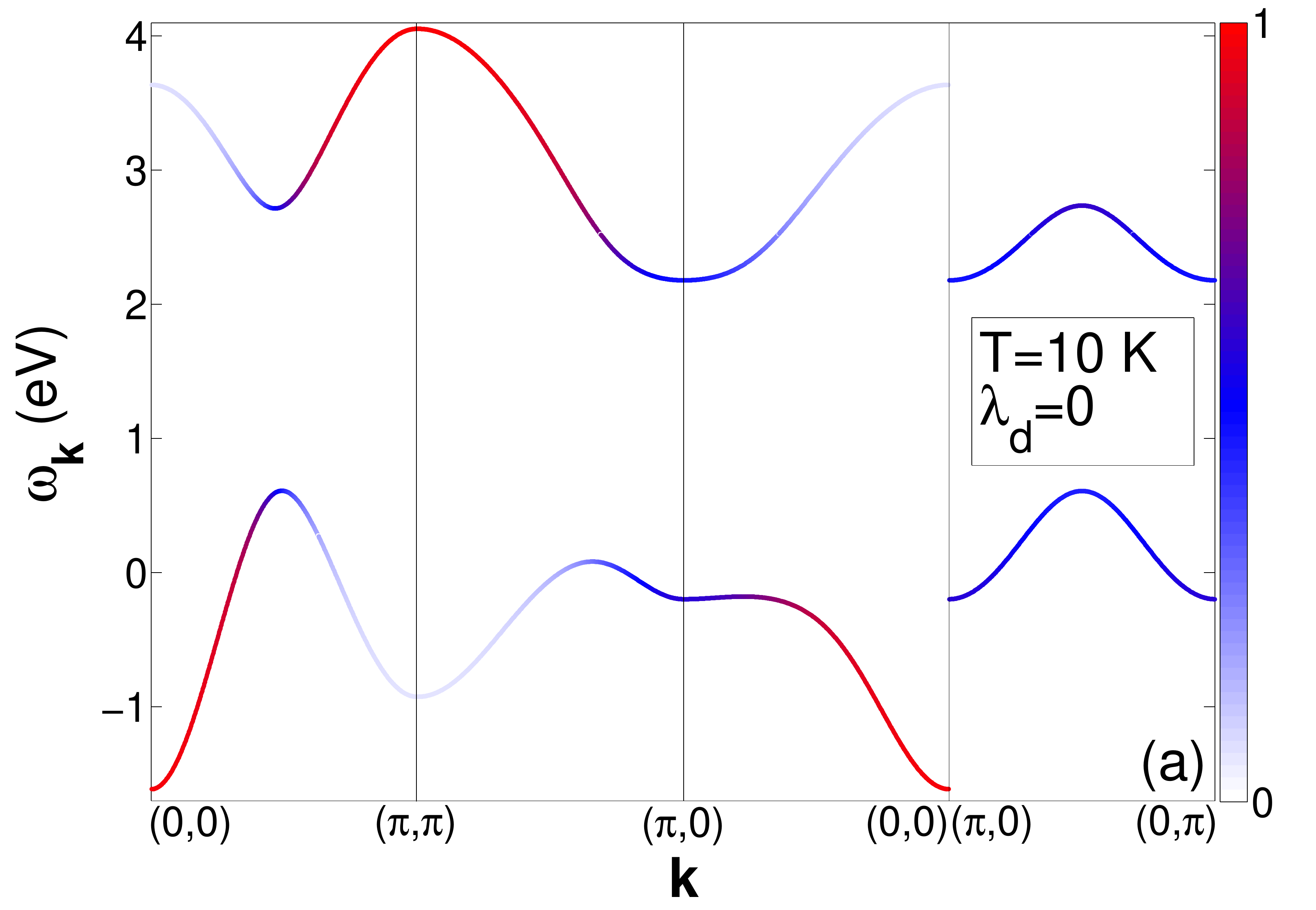}
\includegraphics[width=0.45\linewidth]{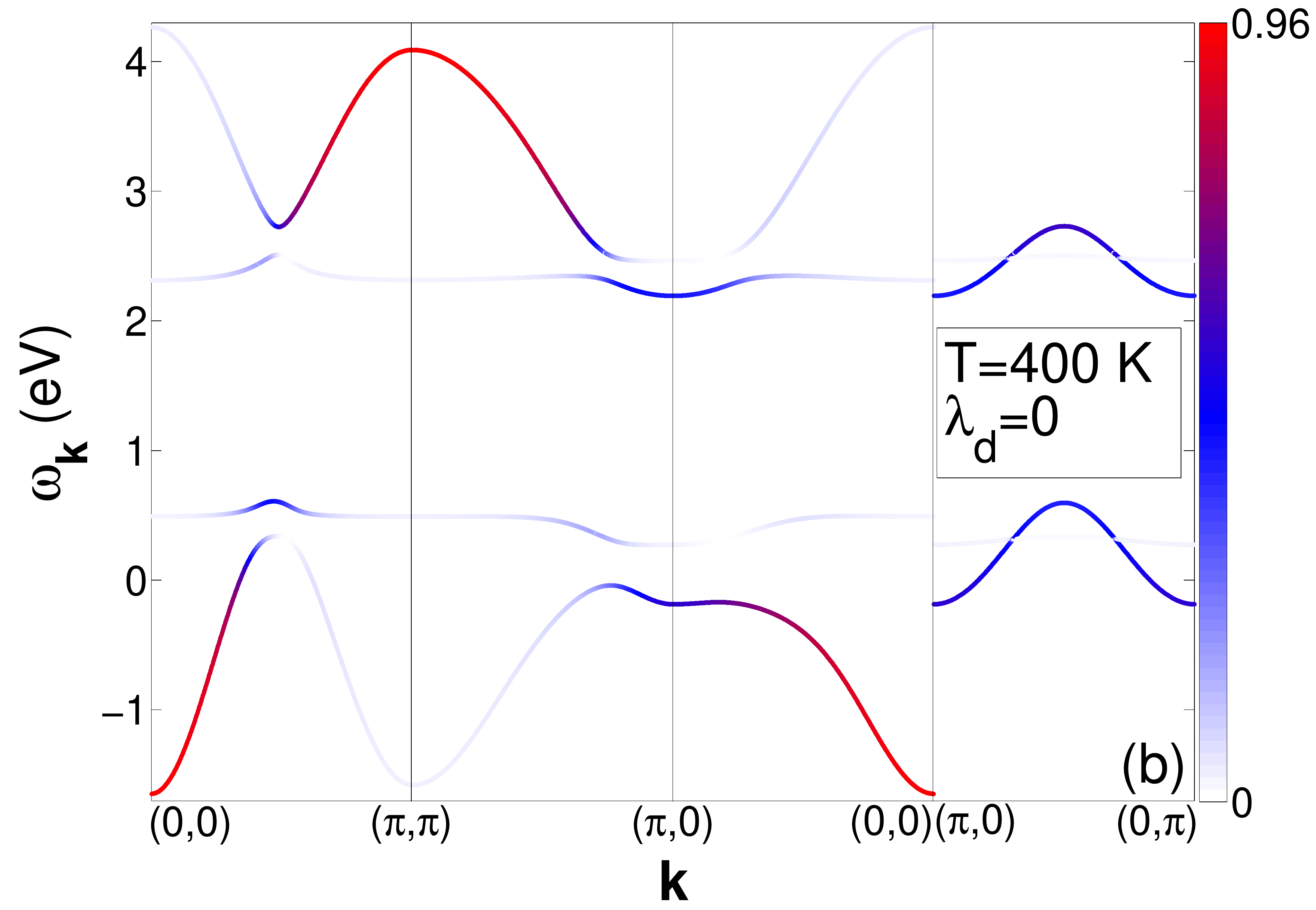}
\includegraphics[width=0.45\linewidth]{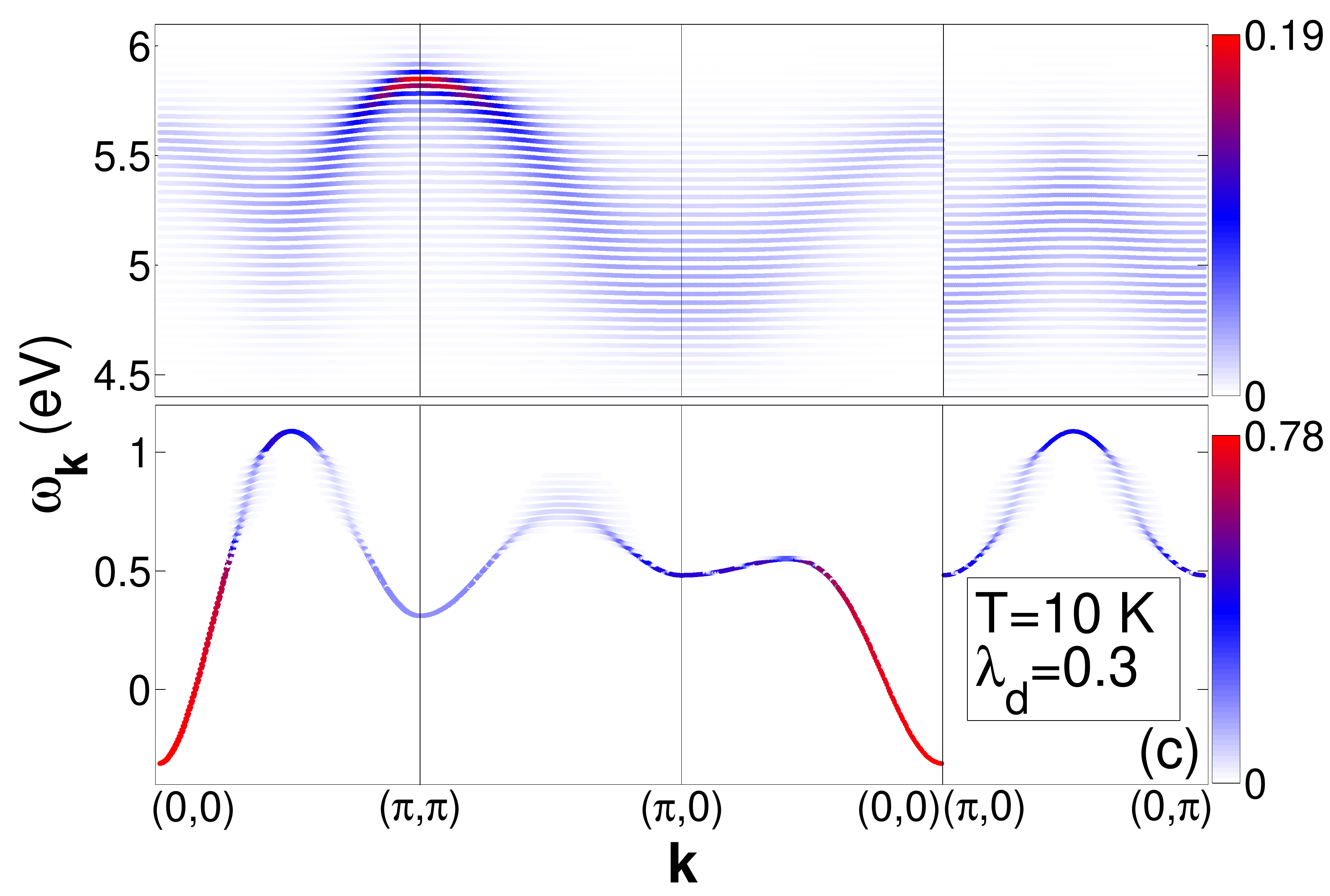}
\includegraphics[width=0.45\linewidth]{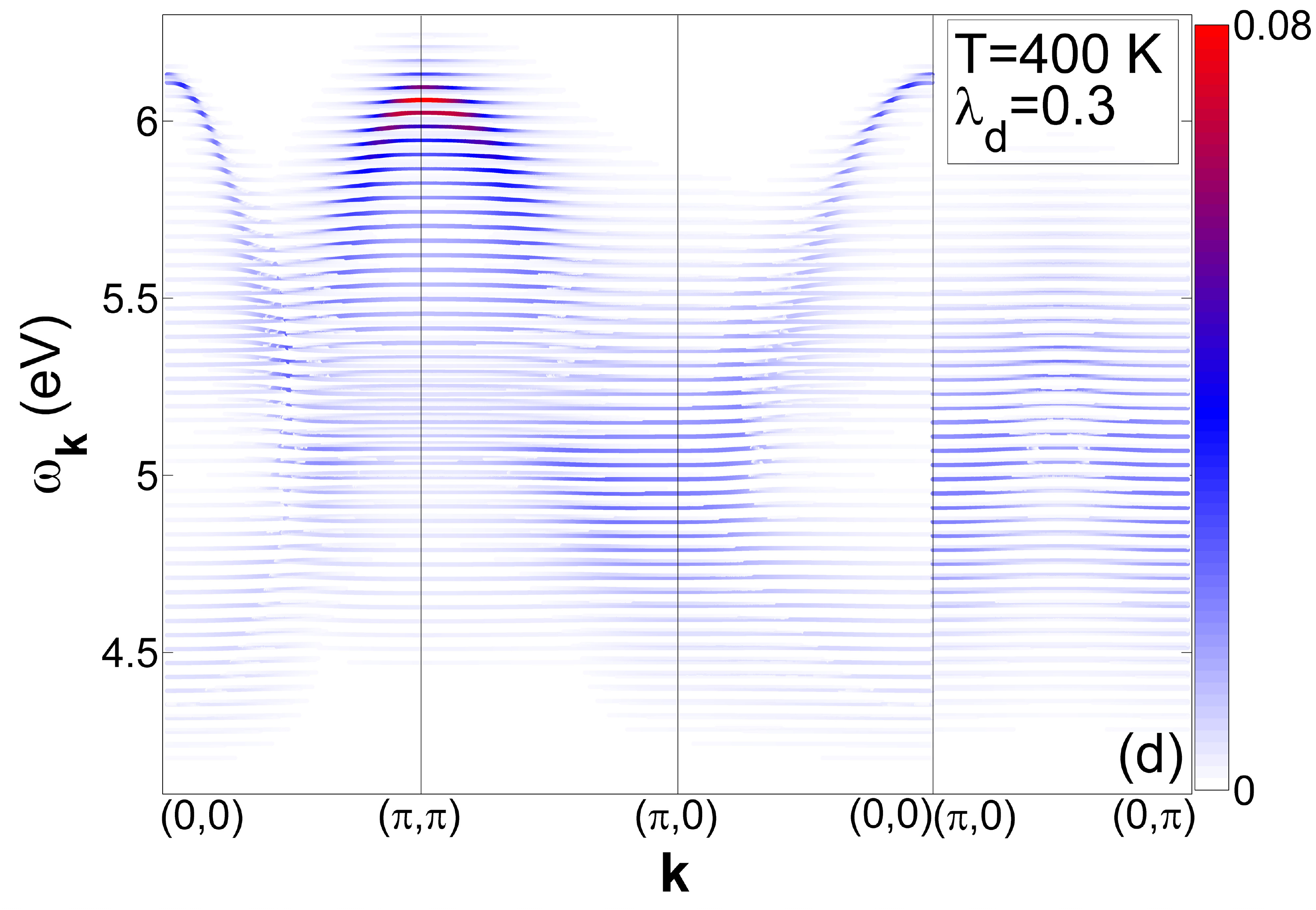}
\includegraphics[width=0.45\linewidth]{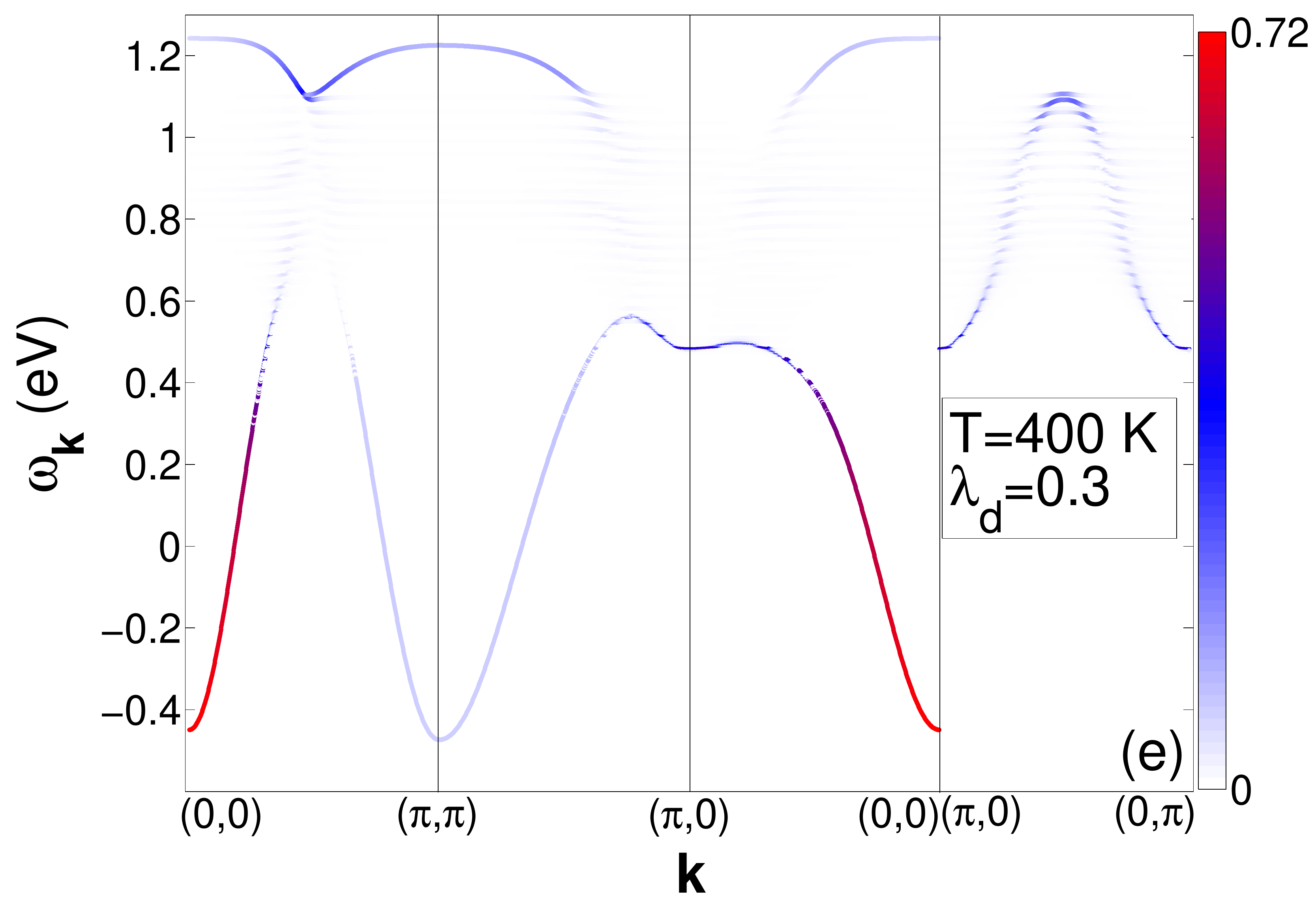}
\caption{\label{fig:bandstr_Tdep_realistic} Evolution of electronic structure with increasing temperature and EPI constant at realistic parameters of electronic system (\ref{realistic_parameters}). Band structure with spectral weight of quasiparticles in each $k$-point (spectral weight is displayed by color) at (a) $T = 10$ K and ${\lambda _d} = 0$, (b) $T = 400$ K and ${\lambda _d} = 0$, (c) $T = 10$ K and ${\lambda _d} = 0.3$, (d),(e) $T = 400$ K and ${\lambda _d} = 0.3$ ((d) conductivity band, (e) valence band).}
\end{figure*}
\begin{figure}
\center
\includegraphics[width=1.0\linewidth]{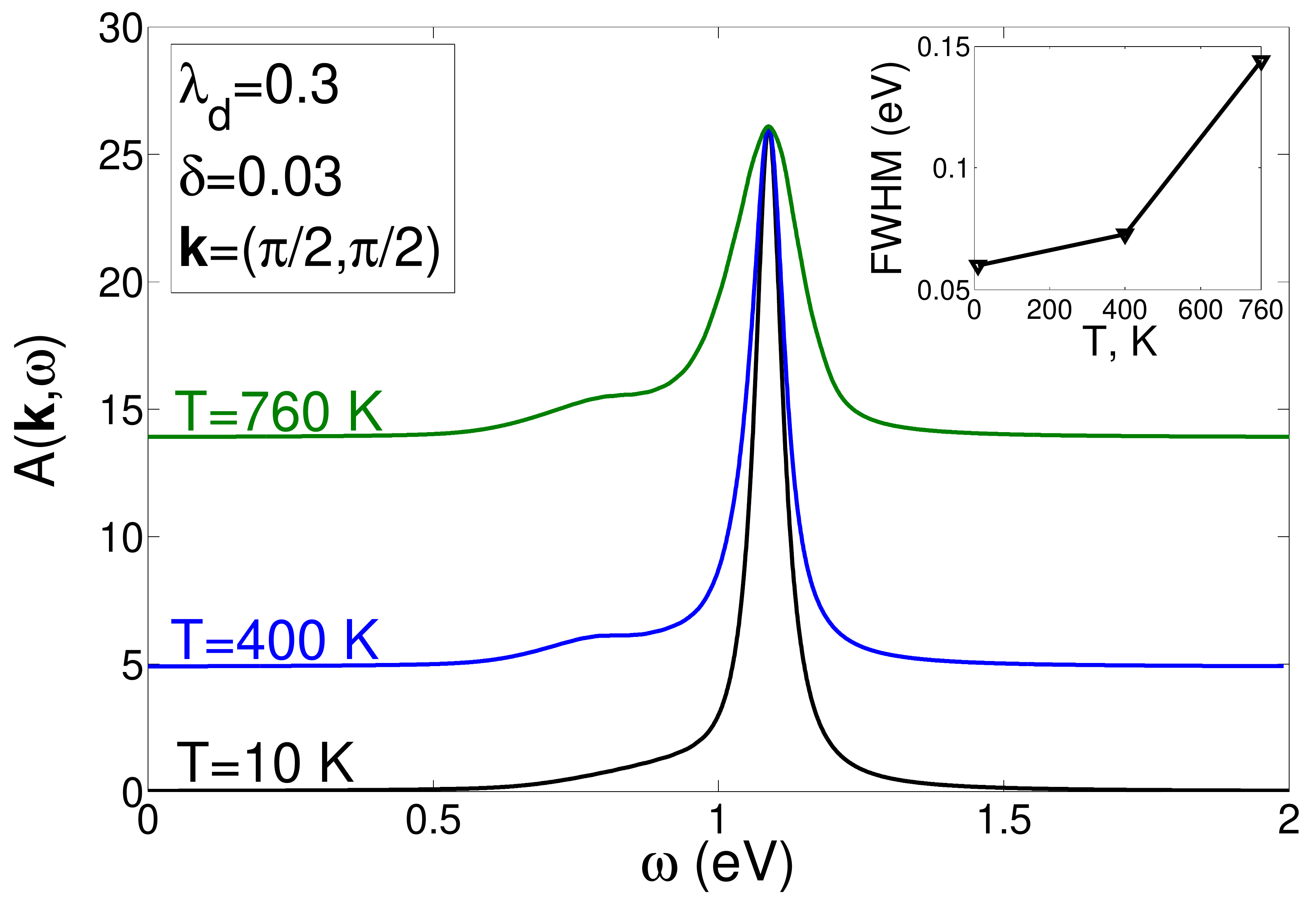}
\caption{\label{fig:SF_Tdep_realistic} The spectral function and its FWHM (inset) at ${\bf{k}} = \left( {\frac{\pi }{2},\frac{\pi }{2}} \right)$ for different temperatures ($T = 10$ K (black line), $T = 400$ K (blue line), $T = 760$ K (green line)) at realistic for La$_2$CuO$_4$ parameters set (\ref{realistic_parameters}). Broadening of spectral function peak with increasing temperature is shown, all curves are rescaled to have the same maximal value.}
\end{figure}

The temperature dependent spectral function is shown in Fig.~\ref{fig:shift_Tdep_withEPI}. The temperature increasing results in the spectral function evolution characterized by appearance of a satellite peak with smaller intensity. Position and intensity of this peak depends on momentum value. High-intensity coherent peak of the electron spectral function in the valence band is shifted with temperature increasing. This shift is caused by reconstruction of quasiparticle band and results from the splitting of $0-0$ band due to hybridization and the weakening of the spin-spin correlations. These factors are momentum depending, i.e. the shift has dispersion. Note ARPES spectra also demonstrate dispersion of the peak shift with increasing temperature (left side of FIG.2 in Ref.~\onlinecite{Kim2002}). At the point ${\bf{k}} = \left( {{\pi  \mathord{\left/
 {\vphantom {\pi  2}} \right.
 \kern-\nulldelimiterspace} 2},{\pi  \mathord{\left/
 {\vphantom {\pi  2}} \right.
 \kern-\nulldelimiterspace} 2}} \right)$ two factors give the opposite contribution and almost compensate each other, the spin-spin correlations decrease wins and the peak of the spectral function demonstrates a small shift deeper into LHB at $T = 760$ K (Fig.~\ref{fig:elstr_Tdep_withEPI}(d) and Fig.~\ref{fig:shift_Tdep_withEPI}). Direction of the shift of the main peak at the points ${\bf{k}} = \left( {{\pi  \mathord{\left/
 {\vphantom {\pi  2}} \right.
 \kern-\nulldelimiterspace} 2},{\pi  \mathord{\left/
 {\vphantom {\pi  2}} \right.
 \kern-\nulldelimiterspace} 2}} \right)$ and ${\bf{k}} = \left( {0.67\pi ,0} \right)$ is in agreement with the peak shift in the ARPES experiments~\cite{Kim2002} at the same points of momentum space. It is seen that at the points ${\bf{k}} = \left( {\pi ,0} \right)$ and ${\bf{k}} = \left( {\pi ,\pi } \right)$ coherent peak shifts to higher electron energies (lower binding energy in ARPES) (Fig.~\ref{fig:shift_Tdep_withEPI}). It results in the decreasing of the insulator gap with increasing temperature. Reflectivity measurements and ${\varepsilon _2}$ spectra in the La$_2$CuO$_4$ show red shift of the peak corresponding to the charge-transfer excitations between the Zhang-Rice singlet and the conductivity band with increasing temperature from $T = 122$ K to $T = 447$ K~\cite{Falck1992,Kastner1998,Choi1999} and this fact also indicate the decreasing of the insulator gap.

\section{Temperature dependence of the electronic structure at realistic parameters of electronic system \label{T_dep_with_EPI_real_par}}

Here we discuss the polaronic band structure for the $p-d$-model parameters
\begin{eqnarray}
{\varepsilon _d} = 0, {\varepsilon _p} = 1.5, {t_{pd}} = 1.36, {t_{pp}} = 0.86 \nonumber \\
{U_d} = 9, {U_p} = 4, {V_{pd}} = 1.5
\label{realistic_parameters}
\end{eqnarray}
calculated for La$_2$CuO$_4$ from \textit{ab-initio} approach~\cite{LDA+GTB} (Fig.~\ref{fig:bandstr_Tdep_realistic}). Comparing the band structure for the over correlated case (\ref{parameters}) and for the realistic set (\ref{realistic_parameters}) one may conclude the following:

I)	In general two band structures shown in Fig.~\ref{fig:bandstr_Tdep_withoutEPI} and Fig.~\ref{fig:bandstr_Tdep_realistic} are similar with the main difference in the value of the insulator gap, the indirect gap ${E_g} = 1.5$ eV for the set (\ref{realistic_parameters}); II) The conductivity band slightly changes its dispersion, for both cases the global minimum takes place at ${\bf{k}}=\left( {\pi ,0} \right)$, while the local minimum at ${\bf{k}}=\left( {{\pi  \mathord{\left/
 {\vphantom {\pi  2}} \right.
 \kern-\nulldelimiterspace} 2},{\pi  \mathord{\left/
 {\vphantom {\pi  2}} \right.
 \kern-\nulldelimiterspace} 2}} \right)$ from Fig.~\ref{fig:bandstr_Tdep_realistic} disappears at Fig.~\ref{fig:bandstr_Tdep_withoutEPI}, the width of the conductivity band is about $2$ eV in Fig.~\ref{fig:bandstr_Tdep_realistic} and about $0.5$ eV in Fig.~\ref{fig:bandstr_Tdep_withoutEPI}; III) Dispersion of the valence band has smaller changes for the both sets of parameters while the bandwidth is also smaller for strongly correlated limit (\ref{parameters}); IV) For strong EPI case ${\lambda _d} = 0.3$ the lost of the polaronic spectral weight in Fig.~\ref{fig:bandstr_withEPI}(a) for the set (\ref{parameters}) is much stronger then for the set (\ref{realistic_parameters}). It means that increasing electron correlations results also in increasing EPI effect on the electronic structure. We relate this finding with the bandwidth decreasing due to stronger electronic correlations. Comparing the EPI effect on the valence and conductivity bands (Fig.~\ref{fig:bandstr_Tdep_realistic}(d),(e)) we would like to emphasize the $10$ times difference in the spectral weight scale, the spectral weight of the polarons from the conductivity band is suppressed much stronger.

At Fig.~\ref{fig:SF_Tdep_realistic} we have shown the temperature dependence of the polaronic spectral weight. Qualitatively the spectral weight and its temperature dependence are similar to the set (\ref{parameters}).

\section{Conclusion \label{Summary}}
Contrary to the standard LDA temperature independent band structure the polaronic band structure of a system with strong electron correlations and strong electron-phonon interaction calculated within our polaronic version of the GTB method is temperature dependent. The proper treatment of strong electron correlations incorporates the spin polaron effects forming the dispersion of the Hubbard fermion at the top of the valence band. The strong electron-phonon interaction results in the formation of the Franck-Condon resonances, that involves the local multiphonon Fermi-type excitations. The hybridization of the Hubbard fermions with the Franck-Condon resonances due to the electron-phonon interaction and occupation of the excited local electronic states with temperature increasing results in the formation of the polaronic band structure. We call these polarons the Hubbard polarons to distinguish them from the standard case of weakly correlated electrons renormalized by the electron-phonon interaction.~\cite{LangFirsov} We have found an interesting interrelation of increasing electron correlations and effects of the electron-phonon interaction. For strong EPI case ${\lambda _d} = 0.3$ the lost of the polaronic spectral weight in the strongly correlated limit is more pronounced than for the realistic more weak Coulomb interaction.

The temperature dependence of electronic spectra calculating in the frameworks of p-GTB method is in qualitative agreement with ARPES spectra. Regime of large EPI was considered to reproduce broad peak of spectral function similar to that observed in the ARPES spectra in the undoped cuprates. We have obtained broadening of the peak of the electron spectral function, the reduction of its maximal intensity, shift of the peak and the decreasing dielectric gap with increasing temperature. The FWHM of the spectral function peak at $400$ K is $1.5$ times larger than at $T = 10$ K and $2.6$ times larger at $760$ K. The peak shift is smaller than in ARPES data.

As concerns a question whether the EPI in cuprates is indeed strong, there is no definite answer today. From theoretical point of view, there is no \textit{ab-initio} EPI calculations revealing large EPI in La$_2$CuO$_4$. But there is no also standard \textit{ab-initio} calculations revealing the insulator ground state of La$_2$CuO$_4$. Only the explicit involvement of strong electron correlations in some hybrid schemes like LDA+DMFT or LDA+GTB results in the correct electronic structure of the charge-transfer insulator. May be future calculations of the EPI in such hybrid schemes will provide the information how strong the EPI in cuprates is. From experimental point of view, the large line width of ARPES in cuprates may be considered as the indication of the large EPI.

\begin{acknowledgments}
Authors are thankful to Russian Science Foundation (project No. 14-12-00061) for financial support.
\end{acknowledgments}

\bibliography{C:/literature_Tdep_pol_ES}

\end{document}